\documentclass[11pt]{article}
\RequirePackage{etex}

\usepackage[T1]{fontenc}
\usepackage[utf8]{inputenc}

\usepackage{fancyhdr}

\newcommand{\jcn}{0}

\usepackage{comment}
\usepackage{booktabs}
\usepackage{nicefrac}
\usepackage{microtype}
\usepackage{lipsum}
\usepackage{graphicx}
\usepackage{multirow}
\usepackage{mathrsfs}
\usepackage[title]{appendix}
\usepackage{xcolor}
\usepackage{textcomp}
\usepackage{manyfoot}
\usepackage{algorithm}
\usepackage{algorithmicx}
\usepackage{algpseudocode}
\usepackage{listings}
\usepackage{float}

\usepackage{natbib}

\usepackage{mathtools}
\mathtoolsset{showonlyrefs}

\usepackage{url}
\usepackage{subcaption}

\usepackage{tikz}
\usetikzlibrary{shapes,decorations,arrows,calc,arrows.meta,fit,positioning,decorations.pathreplacing}
\usepackage{tikz-network}
\usepackage{pgfplots}
\pgfplotsset{compat=1.18}
\usepackage{tikz}
\usepackage{physics}
\usepgfplotslibrary{groupplots}
\usepackage[inline]{enumitem}
\usepackage{multicol}
\usepackage{times}

\usepackage{amsmath,amsfonts,amsthm,amssymb,bm,dsfont}
\usepackage[a4paper,top=2.5cm,bottom=2.5cm,left=2cm,right=2cm]{geometry}
\usepackage{numprint}
\npthousandsep{,}
\npthousandthpartsep{}
\npdecimalsign{.}

\theoremstyle{plain}

\DeclareMathOperator*{\argmin}{argmin}

\newcommand{\titledoc}{Simultaneous global and local clustering in multiplex networks with covariate information}
\newcommand{\titleshort}{Simultaneous global and local clustering in multiplex networks with covariate information}

\providecommand{\keywords}[1]{{\small{\textbf{\textit{Keywords ---}} #1}}}

\usepackage{authblk}

 \author[1]{Joshua Corneck}
\author[1]{Edward A. K. Cohen}
\author[1]{James S. Martin}
\author[2]{Lekha Patel}
\author[2]{Kurtis W. Shuler}
\author[1]{Francesco Sanna Passino}

\affil[1]{Department of Mathematics, Imperial College London, London (United Kingdom)}
\affil[2]{Sandia National Laboratories,
Albuquerque (New Mexico, United States)}
\date{}

\title{\Huge\textbf{\titledoc}}

\allowdisplaybreaks

\usepackage[colorlinks=true,linkcolor=blue,citecolor=blue,urlcolor=blue]{hyperref}
\numberwithin{equation}{section}

\makeatletter
\newenvironment{breakablealgorithm}
  {
   \begin{center}
     \refstepcounter{algorithm}
     \hrule height.8pt depth0pt \kern2pt
     \renewcommand{\caption}[2][\relax]{
       {\raggedright\textbf{\fname@algorithm~\thealgorithm} ##2\par}%
       \ifx\relax##1\relax 
         \addcontentsline{loa}{algorithm}{\protect\numberline{\thealgorithm}##2}%
       \else 
         \addcontentsline{loa}{algorithm}{\protect\numberline{\thealgorithm}##1}%
       \fi
       \kern2pt\hrule\kern2pt
     }
  }{
     \kern2pt\hrule\relax
   \end{center}
  }
\makeatother

\begin{document}

\maketitle

\begin{center}
\textit{This is a pre-print of an article published in the Journal of Complex Networks. The final authenticated version is available online at: \url{https://doi.org/10.1093/comnet/cnaf049}.}
\end{center}

\begin{abstract}
Understanding both global and layer-specific group structures is useful for uncovering complex patterns in networks with multiple interaction types. In this work, we introduce a new model, the hierarchical multiplex stochastic blockmodel (HMPSBM), that simultaneously detects communities within individual layers of a multiplex network while inferring a global node clustering across the layers. A stochastic blockmodel is assumed in each layer, with probabilities of layer-level group memberships determined by a node's global group assignment. Our model uses a Bayesian framework, employing a probit stick-breaking process to construct node-specific mixing proportions over a set of shared Griffiths-Engen-McCloseky (GEM) distributions. These proportions determine layer-level community assignment, allowing for an unknown and varying number of groups across layers, while incorporating nodal covariate information to inform the global clustering. We propose a scalable variational inference procedure with parallelisable updates for application to large networks. Extensive simulation studies demonstrate our model's ability to accurately recover both global and layer-level clusters in complicated settings, and applications to real data showcase the model's effectiveness in uncovering interesting latent network structure.
\end{abstract}


\section{Introduction}\label{sec1}

Modern statistics has seen a growing interest in the use of network models to mathematically describe complex relationships between a large number of connected entities. Network data is often highly complex, with multiple types of connections existing between the \emph{nodes} in the network. To model such heterogeneity of connection types, multiplex networks have been developed as compact representations, with multiple layers being considered simultaneously, each describing different connection types between its nodes. Examples of applications include neuroimaging, where regions of interest in the brain represent nodes and layers correspond to different patients \citep{paul2018}; actions of individuals on a social media platform, where layers might represent ``follow'' or ``retweet'' interactions \citep{greene2013}; or economic data, where layers capture different types of financial exposure \citep{poledna2015}. For a survey of multiplex network applications, see, for example, \cite{kivela2014}.

A task that has garnered much attention in the single-layer network setting is the clustering of nodes into \textit{communities}. One of the most common modelling approaches is the stochastic blockmodel \citep[SBM;][]{Holland}, which models connections between nodes via a latent community structure on a network. In spite of their simplicity, SBMs can be used to consistently estimate and approximate much more complex network structures \citep{airoldi2013}. The classical SBM requires the number of latent communities to be known, but, within a Bayesian framework, this requirement has been circumvented in the literature through a Dirichlet process prior \citep[see, for example,][]{kemp2006, kim2012, amini2024, Corneck2025}. Additionally, the approach of \cite{kim2012} allows for both an unknown number of latent communities, and for the incorporation of nodal metadata.

There has been work to extend single-layer clustering methodology to the multiplex setting, much of which considers variants of a multiplex SBM. Broadly speaking, the approaches can be divided into those that attempt primarily to cluster the layers, and potentially the nodes therein \citep{stanley2015, reyes2016, paul2018, jing2021,  young2022, josephs2023, Zheng2024}, and those that seek only to infer either an across-layer community structure, or within-layer node groups without clustering the layers \citep{amini2024, berlingerio2011, kuncheva2015}. To the best of our knowledge, a methodology that provides both an across-layer, or \textit{global}, clustering of each node alongside a layer-level, or \textit{local}, clustering has not yet been proposed in the literature. In this paper, we aim to address this gap by developing a framework for hierarchical clustering of nodes in a multiplex network that allows heterogeneous clustering across layers, infers latent groups at local and global levels, and incorporates nodal covariate information to guide clustering. This particular type of hierarchical clustering finds relevant applications across several fields. In economics, global clustering can describe trade patterns between entities \citep{fagiolo2010}. In biology, it helps understand how genes or proteins participate in different biological pathways while also inferring overarching functional groupings \citep{barabasi2004}. 

As discussed in \cite{miller2018}, it is important when modelling to understand whether to treat the number of groups is unknown but finite, or as infinite. Broadly speaking, within Bayesian hierarchical modelling, these settings are treated using parametric and nonparametric models, respectively. However, it should be noted that nonparametric models are also frequently employed to tackle the former setting. In the parametric case, an unknown but finite number of groups is often treated by running an inference procedure multiple times and using criteria such as the Akaike information criterion \citep[AIC; ][]{akaike1974} or the Bayesian information criterion \citep[BIC; ][]{gideon1978} to select the optimum number \textit{post hoc}. \cite{green1995} developed reversible-jump MCMC (RJMCMC) as an algorithm that selects the number of groups in a single-pass, but it can be very difficult to implement in practice \citep{miller2018}. Methods have been developed for constructing good reversible jump moves \citep{brooks2003, hastie2012}, but they add high complexity to the model. Priors can also be placed on the number of components to develop more efficient algorithms \cite{miller2018}. To tackle an infinite number of groups, several nonparametric models have been proposed, with two widely used being the hierarchical Dirichlet process \citep[HDP; ][]{teh2006} and the nested Dirichlet process \citep[NDP; ][]{rodriguez_2008}. The infinite nature of these models allows for the selection of the number of latent groups in a single pass. MCMC can be used for these models, broadly using marginalisation \cite{maceachern1994, escobar1995}, truncation \citep{ishwaran2001, ishwaran2002}, or RJMCMC. Variational inference is also popular \cite{blei2006}, adding underestimation of posterior variance at the gain of a large computational speed-up.

With the exception of \cite{reyes2016, josephs2023} and \cite{amini2024}, most existing methods assume a known and finite number of communities for layers and nodes. To the best of our knowledge, there are no existing methodologies that have the flexibility of allowing for both an infinite number (or unknown but finite number) of communities and nodal covariate information simultaneously. Within the network literature, approaches for dealing with an unknown or number of nodes usually rely on Bayesian nonparametric priors; for example, \cite{reyes2016} modify the infinite relational model of \cite{kemp2006} to cluster networks with similar community structure. Similarly, the hierarchical stochastic blockmodel \citep[HSBM;][]{amini2024} employs an HDP to infer layer-level communities whilst sharing information between the layers. The method proposed in this work is closely related to the nested stochastic blockmodel (NSBM) of \cite{josephs2023}, which uses an NDP to provide a clustering of layers and of the nodes within those layers. The nonparametric prior allows for the number of layer communities and communities within layers to be unspecified and automatically selected. However, the NSBM does not provide a global clustering of nodes, nor is it adapted for the inclusion of nodal covariate information, which we address within the model proposed in this work.

The remainder of this article is structured as follows: Section~\ref{sec:model} describes the Bayesian multiplex clustering model used in this work, and the nonparametric priors on which it is based. Section~\ref{sec:inference}
discusses the proposed variational inference procedure for estimating the posteriors of each model parameter. The performance of the proposed inferential procedure is then tested in Section~\ref{sec:sims} on simulated data, and on a real-world trading network dataset in Section~\ref{sec:real_data}, followed by a discussion.

\section{Hierarchical multiplex stochastic blockmodels}
\label{sec:model}

We consider a directed multiplex network with $L$ layers and $N$ nodes. We assume that all layers share the same node set $\mathcal{V} := [N]$, where we write $[m]=\{1,\dots,m\},\ m\in\mathbb N$. Each layer $\ell\in[L]$ consists of a graph $\mathcal{G}_\ell = (\mathcal{V}, \mathcal{E}_\ell)$, where $\mathcal{E}_\ell \subseteq \mathcal{V} \times \mathcal{V}$ is a layer-specific edge set, and we say that $(i,j)\in\mathcal E_\ell$, $i,j\in\mathcal V$, if nodes $i$ and $j$ form a connection of type $\ell$. The full multiplex network is denoted by $\mathcal{G} = (\mathcal{V}, \{\mathcal{E}_1,\dots, \mathcal{E}_L\})$. Each graph $\mathcal{G}_\ell$ could equivalently be represented by its adjacency matrix $A_\ell = \{A_{\ell ij}\}_{i,j=1}^N \in \{0,1\}^{N\times N}$, where $A_{\ell i j} = \mathbb{I}_{\mathcal{E}_\ell}\{(i,j)\}$ and where $\mathbb{I}_{\cdot}\{\cdot\}$ denotes the indicator function. The multiplex network $\mathcal{G}$ can then be represented by an adjacency tensor $A = \{A_{\ell i j},\ \ell \in \{1,\dots, L\},\ i,j \in \{1,\dots,N\}\} \in \{0,1\}^{L\times N\times N}$.

We model the behaviour of each layer by conditionally independent stochastic blockmodels \citep[SBMs;][]{Holland}. To parameterise the SBM of layer $\ell \in [L]$, we assume that each node $i\in\mathcal{V}$ belongs to a group $z_{\ell i} \in \mathbb{N}$. Note that, in a Bayesian nonparametric fashion, we consider the number of clusters to be not fixed a priori, and it is instead treated as a random quantity that can grow with the number of nodes and layers. We model the number of groups in the network as an infinite quantity. Given this layer-level group $z_{\ell i},\ \ell\in[L],\ i\in[N]$, the adjacency matrix $A_\ell$ is modelled as
\begin{equation}
	A_{\ell i j} \mid z_{\ell i}, z_{\ell j}, \rho_{z_{\ell i}z_{\ell j}} \sim \text{Bernoulli}(\rho_{z_{\ell i}z_{\ell j}}),
\end{equation}
where $\rho = \{\rho_{km}\} \in [0,1]^{\mathbb{N} \times \mathbb{N}}$ is a matrix of group connection probabilities. We place beta priors independently upon the entries of $\rho$,
\begin{equation}
	\rho_{km} \sim \text{Beta}(\alpha_0, \beta_0), \quad \text{for all }k,m \in\mathbb{N}.
\end{equation} 
Note that the \emph{same} connectivity probability matrix is used for each layer. This is not a restrictive assumption as we allow for an infinite number of groups in each layer.

\begin{figure}[t]
    \centering
    \includegraphics[width=0.7\textwidth]{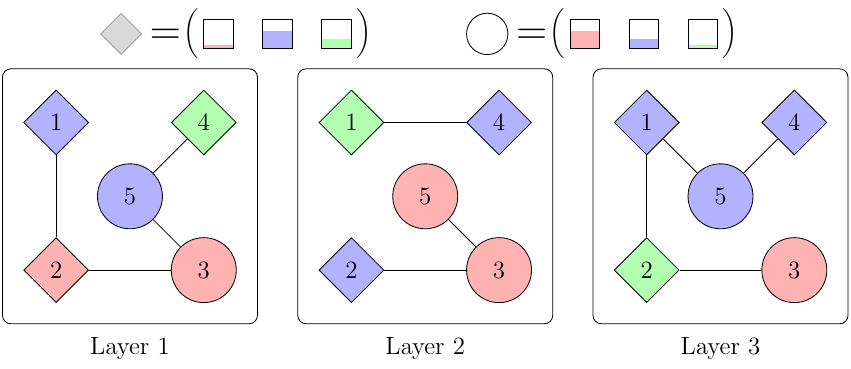}
    \caption{Toy example of the group structure of nodes in a multiplex network with $L=3$ layers corresponding to the illustration in Figure \ref{fig:model_schematic}, with the grey global group as circles and the white as diamonds. There are two global groups, indicated by the shape of the nodes, and three layer groups, indicated by the colours. Probabilities of layer group assignments given the global groups are indicated in the row vectors.}
    \label{fig:toy_network}
\end{figure}

Our key modelling assumption is that subsets of nodes behave similarly across network layers.  To facilitate this assumption in our model, we assign to each node $i \in \mathcal{V}$ a global group $w_i \in \mathbb{N}$, such that
\begin{equation}
    w_i \mid \boldsymbol{\tau}_i \sim \text{Categorical}(\boldsymbol{\tau}_i),\quad i\in[N], \label{eq:wi}
\end{equation}
where $\boldsymbol{\tau}_i$ is a node-specific infinite-dimensional probability vector. Conditional upon $w_i$, we then write 
\begin{equation}
    z_{\ell i} \mid w_i, \boldsymbol{\gamma}_{w_i} \sim \text{Categorical}(\boldsymbol{\gamma}_{w_i}),\quad \ell\in[L],\ i\in[N],
\end{equation}
where $\boldsymbol{\gamma}_k$ is an infinite-dimensional probability vector for each $k \in \mathbb{N}$. We write $\boldsymbol{w} = (w_1,\dots,w_N)^\intercal \in \mathbb{N}^N$ and $z = \{z_{\ell i}\} \in \mathbb{N}^{L \times N}$ for the vector of global group memberships and the matrix of layer group memberships, respectively. We place independent Griffiths-Engen-McCloskey \citep[GEM;][]{pitman2002} prior distributions upon each $\boldsymbol{\gamma}_k$ with parameter $\eta_0$, written $\boldsymbol{\gamma}_k \sim \text{GEM}(\eta_0)$. 
The GEM distribution corresponds to the distribution of proportions obtained under a stick-breaking representation \citep{sethuraman1994} of a Dirichlet process. Therefore, we can express $\boldsymbol{\gamma}_k = (\gamma_{k1}, \gamma_{k2},\dots)^{\intercal}$ in terms of stick-breaking variables $\gamma_{k1}^{\prime},\gamma_{k2}^{\prime},\dots \in [0,1]$ as:
\begin{align*}
	 \gamma_{ks}^{\prime} & \sim \text{Beta}(1,\eta_0), \qquad \qquad \gamma_{ks} = \gamma_{ks}^{\prime} \prod_{r=1}^{s-1}(1 - \gamma_{kr}^{\prime}), 
\end{align*}
and we write $\gamma = \{\gamma_{ks}\}$ and $\gamma^{\prime} = \{\gamma^{\prime}_{ks}\}$ for the matrices storing the parameters. Additionally, we assume that a vector $\boldsymbol{x}_i \in \mathbb{R}^P$ of covariates, which is assumed to be completely observed, is associated to each node $i \in \mathcal{V}$. We use the covariates $\boldsymbol{x}_i$ to construct a the node-specific probability vectors $\boldsymbol{\tau}_i$ in \eqref{eq:wi} for each $i \in \mathcal{V}$. To allow for learning of an unbounded number of communities, we utilise a stick-breaking representation \citep{sethuraman1994} for each $\boldsymbol{\tau}_i$, which ensures that $\sum_{k=1}^\infty \tau_{ik} = 1$, where $\tau_{ik}$ is the probability that node $i$ belongs to global group $k$. 
In particular, we employ the probit stick-breaking construction of \cite{chung_2009} and \cite{rodriguez_2011}. In this model, $\boldsymbol{\tau}_i = (\tau_{i1}, \tau_{i2},\dots)^{\intercal}$ is obtained as 
\begin{equation}
	\tau_{ik} = \Phi\left\{\boldsymbol{x}_i^{\intercal}\boldsymbol{\varphi}_k\right\}\prod_{\ell=1}^{k-1}\left(1 - \Phi\left\{\boldsymbol{x}_i^{\intercal}\boldsymbol{\varphi}_\ell\right\}\right), \quad \text{for each }i \in \mathcal{V}, k \in \mathbb{N},
\end{equation}
where $\Phi\{\cdot\}$ is the cumulative density of a standard normal random variable and $\boldsymbol{\varphi}_k \sim \text{Normal}(\boldsymbol{\varphi}_k^0, \sigma_k^2I_P)$, with $I_P$ the $P \times P$ identity matrix. This approach is similar to \cite{kim2012}, where the authors use a logistic stick-breaking process \citep{ren2011}. We opt for the probit stick-breaking process instead as \cite{rodriguez_2011} prove that an $M$-dimensional truncation of this model converges in total variation norm, and therefore in distribution, to the infinite process. This is particularly useful for Bayesian inference via variational approximations as it demonstrates that truncations can be chosen for arbitrarily accurate inference. We place an independent $\text{Normal}(\boldsymbol{\mu}, I)$ prior distribution on each $\boldsymbol{\varphi}_k^0$, and independent $\text{Inverse\text{-}Gamma}(\nu_0,\omega_0)$ prior distributions on the $\sigma_k^2$, both selected for conjugacy. Write $\varphi = \{\varphi_{kp}\}$, $\varphi^0 = \{\varphi_{kp}^{0}\}$, and $\boldsymbol{\sigma}^2 = (\sigma_1^2,\sigma_2^2,\dots)^\intercal$ to store these parameters as matrices and a vector. In summary, the full model is expressed as follows:
\begin{align}
	A_{\ell i j} \mid z_{\ell i}, z_{\ell j}, \rho_{z_{\ell i}z_{\ell j}} &\sim \text{Bernoulli}(\rho_{z_{\ell i}z_{\ell j}}), & & \text{for all } i, j \in \mathcal{V}, \; i \neq j, \; \ell \in [L], \label{eqn:model_adjacency} \\
	\rho_{km} &\sim \text{Beta}(\alpha_0, \beta_0), & & \text{for all } k,m \in \mathbb{N}, \label{eqn:model_rho} \\
	z_{\ell i} \mid w_i, \boldsymbol{\gamma}_{w_i} &\sim \text{Categorical}(\boldsymbol{\gamma}_{w_i}), & & \text{for all } i \in \mathcal{V}, \; \ell \in [L], \label{eqn:model_z_elli}\\
	w_i \mid \boldsymbol\tau_i &\sim \text{Categorical}(\boldsymbol\tau_i), & & \text{for all } i \in \mathcal{V}, \label{eqn:model_w_i}\\ 
	\tau_{ik} &= \Phi\{\boldsymbol{x}_i^{\intercal}\boldsymbol{\varphi}_k\}\prod_{\ell = 1}^{k-1} \left(1 - \Phi\left\{\boldsymbol{x}_i^{\intercal}\boldsymbol{\varphi}_\ell\right\}\right), & & \text{for all } i \in \mathcal{V},\ k \in \mathbb{N}, \label{eqn:model_tau_ik}\\
	\boldsymbol{\varphi}_k \mid \boldsymbol{\varphi}^0_k, \sigma_k^2 & \sim \text{Normal}(\boldsymbol{\varphi}^0_k,\sigma_k^2I_P), & & \text{for all } k \in \mathbb{N}, \label{eqn:model_varphi}\\
	\boldsymbol{\varphi}^0_k &\sim \text{Normal}(\boldsymbol{\mu}, I_P), & & \text{for all } k \in \mathbb{N},
	\label{eqn:model_mu} \\
	\sigma_k^2 &\sim \text{Inverse\text{-}Gamma}(\nu_0, \omega_0), & & \text{for all } k \in \mathbb{N}\label{eqn:model_sigma}, \\
	\boldsymbol\gamma_k &\sim \text{GEM}(\eta_0), & & \text{for all } k \in \mathbb{N}. \label{eqn:model_gamma_k}
\end{align}
We call the proposed model \emph{hierarchical multiplex stochastic blockmodel (HMPSBM)}. Figure \ref{fig:model_schematic} provides an illustration of the parameters in the model, and Figure \ref{fig:model_DAG} provides a representation of the HMPSBM as a directed acyclic graph (DAG). Figure \ref{fig:toy_network} shows an illustration of the global and layer-level groups in a small network with 3 layers.\\

Estimating the number of components in a finite mixture model presents consistency challenges under model misspecification. These issues arise both in finite parametric frameworks \cite{cai2011} and when using nonparametric priors \cite{miller2013, miller2014}, unless the Dirichlet process concentration parameter is properly specified \cite{ascolani2022}. Therefore, in practical applications, perfect recovery of the true number of components from the data generating process should not be expected. However, simulation studies in Section \ref{sec:sims} show that, under the HMPSBM, difficulties in estimating the correct number of latent groups did not arise.\\
\if1\jcn
{
\begin{figure}[t]
    \centering
    \subfigure[Illustration of model parameters $\boldsymbol{\gamma}_k,\ \boldsymbol{\varphi}_k,\ \boldsymbol{\tau}_i,\ \boldsymbol{w}$ and $\boldsymbol{z}_\ell$ for an HMPSBM on $N=5$ nodes with $L=3$, 2 global groups and 3 layer-level groups in each layer.]{
        \includegraphics[width=0.415\textwidth]{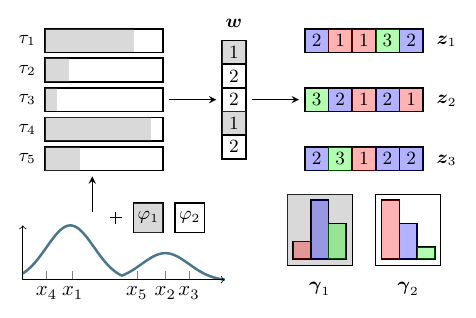}
        \label{fig:model_schematic}
    }
    \hspace{5mm}
    \subfigure[A directed acyclic graph for the HMPSBM model.]{
        \includegraphics[width=0.485\textwidth]{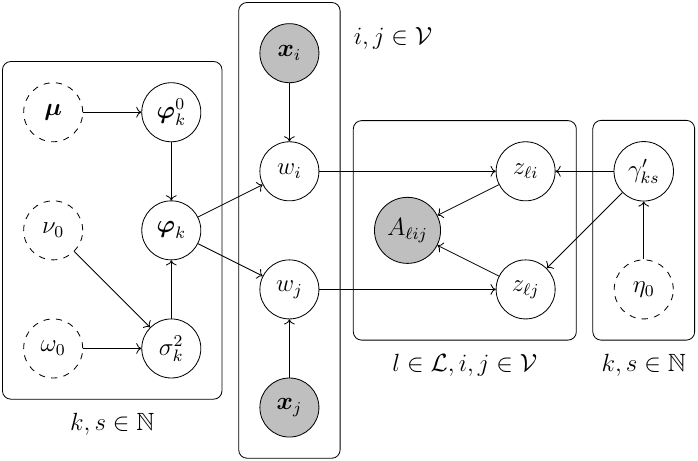}
        \label{fig:model_DAG}
    }
    \caption{Illustration of the parameters and directed acyclic graph for the HMPSBM in \eqref{eqn:model_adjacency}-\eqref{eqn:model_gamma_k}.}
\end{figure}
}\else{
\begin{figure}[t]
    \centering
    \begin{subfigure}[b]{0.415\textwidth}
        \centering \includegraphics[width=\linewidth]{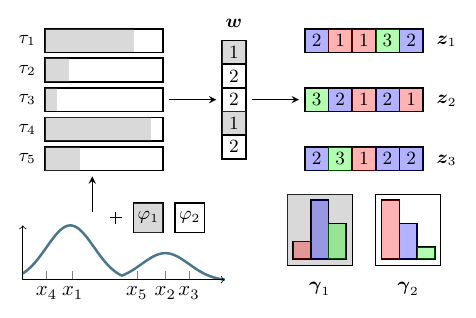}
    \caption{Illustration of model parameters $\boldsymbol{\gamma}_k,\ \boldsymbol{\varphi}_k,\ \boldsymbol{\tau}_i,\ \boldsymbol{w}$ and $\boldsymbol{z}_\ell$ for an HMPSBM on 5 nodes and 3 layers, with 2 global groups and 3 layer-level groups in each layer.}
    \label{fig:model_schematic}
    \end{subfigure}
    \hspace{5mm}
    \begin{subfigure}[b]{0.485\textwidth}
        \centering \includegraphics[width=\linewidth]{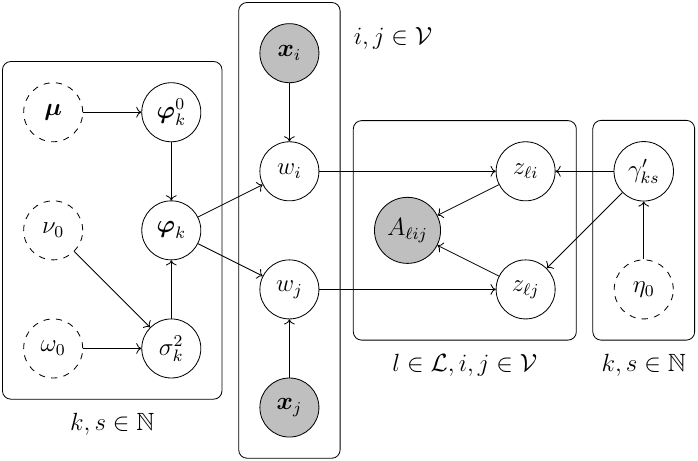}
    \caption{A directed acyclic graph for the HMPSBM model.}
    \label{fig:model_DAG}
    \end{subfigure}
    \caption{Illustration of the parameters and directed acyclic graph for the HMPSBM in \eqref{eqn:model_adjacency}-\eqref{eqn:model_gamma_k}.}
\end{figure}
}\fi

\noindent 




The proposed model shares similarities with the hierarchical stochastic blockmodel (HSBM) proposed in \cite{amini2024} and the nested stochastic blockmodel (NSBM) in \cite{josephs2023}. Through the HSBM, \cite{amini2024} seek to cluster nodes within layers of a multiplex network. A random partition prior is placed upon the node communities in each layer which is based upon the hierarchical Dirichlet process \citep[HDP;][]{teh2006}. The choice of an HDP hierarchy is to allow for sharing of information across layers, rather than treating them as independent. Conditional upon the group memberships of the nodes in a layer, an SBM is placed upon the connections. The connection probability matrix is shared between layers, which ensures that layer groups correspond to one another.

On the other hand, the NSBM is used in \cite{josephs2023} for clustering of layers in a multiplex network where each layer follows an SBM. Layers are assigned to a community which determines its connection probability matrix. In the NSBM, the labels for the layers and layer-level groups are sampled from a model consisting of a label-only NDP coupled with independent draws from a beta distribution to form the group-connection probabilities in each layer. This corresponds to drawing $L$ times, once for each layer, from a mixture of GEM distributions 
with atoms and masses sampled from GEM distributions. 
This mixture model determines the community $z_\ell \in \mathbb{N}$ of layer $\ell$, and assigns to it a class-specific GEM distribution from which the node communities $\xi_{i \ell}$ of each node $i$ in that layer are drawn. A collection of connectivity parameters $\{\eta_{xyz_\ell}\}_{x,y\in\mathbb{N}}$ for the layer-community SBM is then associated with each layer, and a connection between nodes $i$ and $j$ in layer $\ell$ is then formed with probability $\eta_{\xi_{i\ell}\xi_{j\ell}z_\ell}$.
The key distinction between our work and \cite{amini2024} and \cite{josephs2023} is the goal of obtaining a global clustering of each node. Furthermore, we develop a nonparametric framework that allows for the incorporation of covariate information. In our model, we allow for node-specific mixtures of the \textit{same} GEM distributions, with the mixture probabilities $\boldsymbol{\tau}_i$ driven by nodal covariates. That is, the atoms of each mixture component are the same, but the mixture probabilities differ, being informed by the features. This choice ensures that the clustering of each nodes is comparable and so we obtain a global grouping.

\section{Inference via variational Bayesian methods}
\label{sec:inference}
\label{subsec:variational_bayes}

The main objective of inference on the HMPSBM model is to evaluate the posterior distribution of all model parameters, given the adjacency tensor $A$ and the design matrix of covariates $X = \{\boldsymbol{x}_i^{\intercal}\}_{i=1}^N \in \mathbb{R}^{N \times P}$. 
The full posterior distribution of the HMPSBM model in \eqref{eqn:model_adjacency}--\eqref{eqn:model_gamma_k} factorises as
\begin{multline}
    p(\rho, \boldsymbol{w}, z, \varphi, \varphi^0, \boldsymbol{\sigma}^2, \gamma^{\prime} \mid A, X) \propto \\
    p(A \mid z, \rho)\ p(z \mid \boldsymbol{w},\gamma^{\prime})\ p(\boldsymbol{w} \mid \varphi, X)\ p(\varphi \mid \varphi^0, \boldsymbol{\sigma}^2)\ p(\gamma^{\prime})\ p(\rho)\  p(\varphi^0)\ p(\boldsymbol{\sigma}^2).
	\label{eqn:joint_density}
\end{multline}
The normalising constant for the distribution in \eqref{eqn:joint_density} is not available in closed form, therefore it is necessary to resort to sampling or approximation methods to study the posterior distribution. In particular, we use variational Bayesian (VB) inference
to obtain an analytic approximation to the posterior distribution 
\citep{wang2019}. VB transforms an inference problem into an optimisation problem, which allows for parallelisation and mini-batching \citep{blei-2017}, making it more scalable to large datasets than Markov Chain Monte Carlo (MCMC). In a VB approach, a family of distributions $\mathcal F$ is posited on the parameter space of $\Theta = \{\rho,\gamma^{\prime}, \varphi, \varphi^0, \boldsymbol{\sigma}^2, z, \boldsymbol{w}\}$, which is partitioned into $S$ disjoint subsets,
and one seeks to select the component $q^\ast(\Theta)\in \mathcal F$ that is closest to the true posterior $p(\Theta \mid X, A)$ in the sense of the Kullback-Leibler (KL) divergence. A mathematically convenient choice for $\mathcal F$ is the mean-field variational family:
\begin{equation}
	\mathcal{F} = \left\{q:q(\Theta) = \prod_{s=1}^S q_s(\Theta_s)\right\},
\end{equation} 
which is the set of distributions that factorise over the parameters. The approximating distribution $q^*(\Theta)\in\mathcal{F}$ is found by maximising the evidence lower bound (ELBO)
\begin{equation}
    \mathcal L(q)=\text{ELBO}(q) = \mathbb{E}_q\left\{\log p(\Theta, X, A\right)\} - \mathbb{E}_q\left\{\log q(\Theta)\right\},
\end{equation}
which, in practice, is done by solving the equivalent problem:
\begin{equation}
	q^\ast(\Theta) = \argmin_{q \in \mathcal{F}} \text{KL}\left[q(\Theta)\ \| \ p(\Theta|X, A)\right].
	\label{eqn:VB_objective}
\end{equation}
Despite the simplicity of $\mathcal{F}$, the solution $q^\ast(\Theta)$ to \eqref{eqn:VB_objective} is usually not available analytically. A common solution is to approximate $q^\ast(\Theta)$ by another factorisable distribution $\hat{q}(\Theta) = \prod_{s=1}^S \hat{q}(\Theta_s)$ using the Coordinate Ascent Variational Inference (CAVI) algorithm \citep{bishop-2006, blei-2017}. CAVI iteratively updates each component of $\hat{q}(\Theta)$ while keeping the others fixed:
\begin{equation}
	\hat{q}_{s}^{(t+1)}(\Theta_s) = \argmin_{q_{s}} \text{KL}\left[q_{s}(\Theta_s) \prod_{r<s}\hat{q}_{r}^{(t+1)}(\Theta_r) \prod_{r>s}\hat{q}_{r}^{(t)}(\Theta_r)\ \Bigg|\Bigg|\ p(\Theta|X, A)\right],
	\label{eqn:cavi_target}
\end{equation}
for $s \in \{1,\dots,S\},\ t=0,1,2,\dots$
until some convergence criterion is met. 
Under the mean-field variational approximation, \eqref{eqn:cavi_target} has the following solution \citep{bishop-2006}:
\begin{equation}
	\hat{q}_{s}(\Theta_s) \propto \exp\{\mathbb{E}_{\Theta_{-s}}\left[\log p(\Theta_s \mid \Theta_{-s}, X, A)\right]\} \propto \exp\{\mathbb{E}_{-\Theta_s}\left[\log p(\Theta,X, A)\right]\},
	\label{eq:cavi_update_sol}
\end{equation}
where the notation $\mathbb{E}_{\Theta_{-s}}\{\cdot\}$ is used to denote the expectation with respect to all components of $\Theta$ except $\Theta_s$ \citep{blei-2017}. In the case of a conjugate model setting, we are able to obtain each $p(\Theta_s \mid \Theta_{-s}, X, A)$ in closed-form and so can compute the optimal update in \eqref{eq:cavi_update_sol} exactly. If a conjugate prior for $\Theta_s$ is not analytically available, we must instead make a choice for a parameterised variational family depending on parameter vector $\Lambda_s$, written 
$q_s(\Theta_s) = f_s(\Theta_s; \Lambda_s)$. Then, at update $t$, the optimal choice of $\hat{q}^{(t)}_s(\Theta_s)$ is to select $\hat{\Lambda}^{(t)}_s$ by maximising the ELBO with respect to $\Lambda_s$, holding all other parameters fixed, and then to set  $\hat{q}^{(t)}_s(\Theta_s) = f_s(\Theta_s; \hat{\Lambda}^{(t)}_s)$.

It should be noted that CAVI is only guaranteed to achieve a local minimum, and hence is sensitive to intialisation \citep{Zhang2020-re, blei-2017}. Furthermore, updates in \eqref{eq:cavi_update_sol} have an explicit ordering. The ordering of the updates can also have implications on the convergence properties of the algorithm \citep{Ray22}, similarly to Gibbs sampling \citep[see, for example,][]{vanDyk08}.

MCMC can be used as an alternative to target the posterior in Dirichlet process models. Broadly, approaches fall into three categories: methods that marginalise out infinite-dimensional parameters \citep{maceachern1994, escobar1995}; methods that approximate via truncation \citep{ishwaran2001, ishwaran2002}; and reversible-jump MCMC (RJMCMC) methods \citep{green2001, dahl2003, jain2004}. Marginalisation relies on the Pólya urn scheme, which, as discussed in \cite{rodriguez_2008}, becomes problematic in the nested setting due to the need to compute complex predictive distributions. While RJMCMC could in principle be applied, constructing efficient trans-dimensional proposals for nested structures is challenging; as shown by \cite{rannala2013}, split moves that introduce many parameters jointly can result in low acceptance rates and poor mixing. For the HDP, \cite{amini2019} developed an exact slice sampler, later applied in \cite{amini2024}, but no such analogue is available for the nested case. Truncation-based methods are possible, with \cite{josephs2023} considering four different Gibbs samplers, but with our addition of covariates and no efficient sampling scheme, computation time quickly becomes a limiting factor. We therefore adopt variational inference in this work.

\subsection{Variational approximation of the posterior distribution for the HMPSBM}
\label{subsec:variational_approxs}

In order to perform variational Bayesian inference on the HMPSBM model, we choose a mean-field approximation to the true posterior \eqref{eqn:joint_density}, which takes the following product form:
\begin{align}
	q(\boldsymbol{w}, \boldsymbol{z}, \boldsymbol{\rho}, \boldsymbol{\varphi}, \boldsymbol{\varphi}^0,\boldsymbol{\sigma}^2, \boldsymbol{\gamma}^{\prime}) =& \prod_{i\in \mathcal{V}} q(w_i) \times \prod_{\ell \in [L]}\prod_{i\in\mathcal{V}} q(z_{\ell i}) \times \prod_{k,m=1}^{M_z} q(\rho_{km}) \times \prod_{k=1}^{M_w} q(\boldsymbol{\varphi}_k) \\
    &\times \prod_{k=1}^{M_w} q(\boldsymbol{\varphi}_k^0) \times \prod_{k=1}^{M_w} q(\sigma_k^2) \times \prod_{k=1}^{M_w}\prod_{s=1}^{M_z} q(\gamma_{ks}^{\prime}).
\end{align}
A variational approximation is posited directly upon the stick-breaking variables of the GEM distribution, following the approach of \cite{blei2006}, and two truncation parameters $M_z, M_w \in \mathbb{N}$ are introduced, implying that $q(\rho_{km}) = \delta_0(\rho_{km})$ for $\{k,m \in \mathbb{N} : \max\{k,m\} > M_z\}$, $q(\boldsymbol{\varphi}_k) = \delta_0(\boldsymbol{\varphi}_k)$ for $\{k\in\mathbb{N} : k > M_w\}$, $q(\boldsymbol{\varphi}_k^0) = \delta_0(\boldsymbol{\varphi}_k^0)$ for $\{k\in\mathbb{N} : k > M_w\}$, $q(\sigma_k^2) = \delta_0(\sigma_k^2)$ for $\{k\in\mathbb{N} : k > M_w\}$, and $q(\gamma_{ks}^{\prime}) = \delta_0(\gamma_{ks}^{\prime})$ for $\{k,s \in \mathbb{N}: k > M_w \mid m > M_z\}$. Here we define $\delta_0(\cdot)$ to be the Dirac delta function at 0. Note that the model set-up of \eqref{eqn:model_adjacency}-\eqref{eqn:model_gamma_k} is infinite-dimensional, and we only make a finite approximation to it in the variational procedure.

All complete conditionals take the form of standard distributions, except for those of $\{\boldsymbol{\varphi}_{k}\}_{k = 1}^{M_w}$ (see Appendix \ref{app:complete_conditional}). For variables with closed-form conditional distributions, we choose the corresponding approximating distribution to have the same form of the complete conditional for the corresponding component and provide explicit formulae for their parameters in \eqref{eqn:variational_w}-\eqref{eqn:variational_beta_rho}. Under the variational approximation $q(\Theta)$, $w_i$ and $z_{\ell i}$ have categorical distributions with parameters $\Tilde{\boldsymbol{\phi}}_{w_i}$ and $\Tilde{\boldsymbol{\phi}}_{z_{\ell i}}$. The variational approximations for the parameters $\rho_{km}$ and $\gamma_{ks}^{\prime}$ are taken to be Beta distributions with shape parameters $(\Tilde{\alpha}_{\rho_{km}},\Tilde{\beta}_{\rho_{km}})$ and $(\Tilde{\alpha}_{\gamma_{ks}},\Tilde{\beta}_{\gamma_{ks}})$ respectively. 
The parameters $\boldsymbol{\varphi}_k^0$ are approximated as multivariate normal random variables, with means $\Tilde{\boldsymbol{\theta}}_{\varphi_k^0} \in \mathbb{R}^P$ and covariance matrices $\widetilde{\Sigma}_{\varphi_k^0} \in \mathbb{R}^{P \times P}$, whereas $\sigma_k^2$ is taken to have an inverse-gamma variational approximation with shape and scale $\Tilde\nu_k$ and $\Tilde\omega_k$.

For $\boldsymbol{\varphi}_k$, which does not have complete conditional in standard form, we propose to approximate its distributions by a multivariate normal distribution with mean $\Tilde{\boldsymbol{\theta}}_{\varphi_k}$ and covariance $\widetilde{\Sigma}_{\varphi_k}$. Gradient ascent steps are then taken using the Adam algorithm of \cite{king2015} to maximise the ELBO with respect to $\boldsymbol{\varphi}_k$ while holding all other parameters fixed. 

Using these forms, Appendices \ref{app:complete_conditional} and \ref{app:gradients} show that the optimal updates take the form:
\begin{enumerate}
    \item $\hat{q}(w_i) = \text{Categorical}\left(\Tilde{\boldsymbol{\phi}}_{w_i}\right)$ for all $i \in \mathcal{V}$, where $\Tilde{\boldsymbol{\phi}}_{w_i}=(\Tilde{{\phi}}_{w_i,1}, \dots, \Tilde{{\phi}}_{w_i,M_w})$ is defined as:
    \begin{align}
    \Tilde{\boldsymbol{\phi}}_{w_i,w} \propto \exp\Bigg\{&\sum_{\ell \in [L]} \sum_{k=1}^{M_z} \boldsymbol{\phi}_{z_{\ell i},k}\left(\psi(\Tilde{\alpha}_{\gamma_{wk}}) - \psi(\Tilde{\alpha}_{\gamma_{wk}} + \Tilde{\beta}_{\gamma_{wk}}) + \sum_{s=1}^{k-1}\left[\psi(\Tilde{\beta}_{\gamma_{ws}}) - \psi(\Tilde{\alpha}_{\gamma_{ws}} + \Tilde{\beta}_{\gamma_{ws}})\right]\right)\\
    &+ \mathbb{E}_q\left[\log \tau_{iw}\right]\Bigg\},
    \label{eqn:variational_w}
    \end{align}
where $\psi(\cdot)$ denotes the digamma function.
\item $\hat{q}(z_{\ell i}) = \text{Categorical}\left(\Tilde{\boldsymbol{\phi}}_{z_{\ell i}}\right)$ for all $\ell \in [L]$ and $i \in \mathcal{V}$, where $\Tilde{\boldsymbol{\phi}}_{z_{\ell i}} =(\Tilde{{\phi}}_{z_{\ell i},1}, \dots, \Tilde{{\phi}}_{z_{\ell i},M_z})$ is defined as:
\begin{align}
	\Tilde{\boldsymbol{\phi}}_{z_{\ell i},k} \propto \exp\Bigg\{&\sum_{w=1}^{M_w} \Tilde{\boldsymbol{\phi}}_{w_{i},w} \left(\psi(\Tilde{\alpha}_{\gamma_{wk}}) - \psi(\Tilde{\alpha}_{\gamma_{wk}} + \Tilde{\beta}_{\gamma_{wk}}) + \sum_{s=1}^{k-1}\left[\psi(\Tilde{\beta}_{\gamma_{ws}}) - \psi(\Tilde{\alpha}_{\gamma_{ws}} + \Tilde{\beta}_{\gamma_{ws}})\right]\right)\\ 
    &+  \sum_{j\in\mathcal{V}\setminus\{i\}} \sum_{s=1}^{M_z} \Tilde{\boldsymbol{\phi}}_{z_{\ell j},s}\bigg[A_{\ell i j}\left(\psi(\Tilde{\alpha}_{\rho_{ks}}) - \psi(\Tilde{\alpha}_{\rho_{ks}} + \Tilde{\beta}_{\rho_{ks}})\right) \\
    &+ (1 - A_{\ell ij})\left(\psi(\Tilde{\beta}_{\rho_{ks}}) - \psi(\Tilde{\alpha}_{\rho_{ks}} + \Tilde{\beta}_{\rho_{ks}})\right)\\ 
	& + A_{\ell ji}\left(\psi(\Tilde{\alpha}_{\rho_{sk}}) - \psi(\Tilde{\alpha}_{\rho_{sk}} + \Tilde{\beta}_{\rho_{sk}})\right) \\
    &+ (1 - A_{\ell ji})\left(\psi(\Tilde{\beta}_{\rho_{sk}}) - \psi(\Tilde{\alpha}_{\rho_{sk}} + \Tilde{\beta}_{\rho_{sk}})\right)\bigg]\Bigg\},
    \label{eqn:variational_z}
\end{align}
\item $\hat{q}(\boldsymbol{\varphi}_k) = \text{Normal}\left(\Tilde{\boldsymbol{\theta}}_{\varphi_k}, \widetilde{\Sigma}_{\varphi_k}\right)$ for $k=1,\dots, M_w$. The parameters $\Tilde{\boldsymbol{\theta}}_{\varphi_k}$ and $\widetilde{\Sigma}_{\varphi_k}$ are updated using the following gradients of the ELBO:
\begin{align}
    \pdv{\mathcal{L}}{\Tilde{\boldsymbol{\theta}}_{\varphi_k}} &= \sum_{i=1}^N \Tilde{\boldsymbol{\phi}}_{w_i,k}\mathbb{E}_q\left\{\widetilde{\Sigma}_{\varphi_k}^{-1}(\boldsymbol{\varphi}_k - \Tilde{\boldsymbol{\theta}}_{\varphi_k}) \log\left[\Phi\left(\boldsymbol{x}_i^\intercal\boldsymbol{\varphi}_k\right)\right]\right\}\\ 
    &+ \sum_{i=1}^N \sum_{m=k+1}^\infty \Tilde{\boldsymbol{\phi}}_{w_i,m}\mathbb{E}_q\left\{\widetilde{\Sigma}_{\varphi_k}^{-1}(\boldsymbol{\varphi}_k - \Tilde{\boldsymbol{\theta}}_{\varphi_k}) \log\left[1 - \Phi\left(\boldsymbol{x}_i^\intercal\boldsymbol{\varphi}_k\right)\right]\right\} - \frac{\Tilde{\nu}_k}{\Tilde{\omega}_k}\left(\Tilde{\boldsymbol{\theta}}_{\varphi_k} - \Tilde{\boldsymbol{\theta}}_{\varphi_k^0}\right) \label{eqn:variational_theta}\\
    \pdv{\mathcal{L}}{\widetilde{\Sigma}_{\varphi_k}} &= \sum_{i=1}^N \Tilde{\boldsymbol{\phi}}_{w_i,k}\mathbb{E}_q\left\{\frac{1}{2}\left(\widetilde{\Sigma}_{\varphi_k}^{-1}(\boldsymbol{\varphi}_k - \Tilde{\boldsymbol{\theta}}_{\varphi_k})(\boldsymbol{\varphi}_k - \Tilde{\boldsymbol{\theta}}_{\varphi_k})^\intercal\widetilde{\Sigma}_{\varphi_k}^{-1} -\widetilde{\Sigma}_{\varphi_k}^{-1}\right)\log\left[\Phi\left(\boldsymbol{x}_i^\intercal\boldsymbol{\varphi}_k\right)\right]\right\}\\ 
    &+ \sum_{i=1}^N \sum_{m=k+1}^\infty \Tilde{\boldsymbol{\phi}}_{w_i,m}\mathbb{E}_q\left\{\frac{1}{2}\left(\widetilde{\Sigma}_{\varphi_k}^{-1}(\boldsymbol{\varphi}_k - \Tilde{\boldsymbol{\theta}}_{\varphi_k})(\boldsymbol{\varphi}_k - \Tilde{\boldsymbol{\theta}}_{\varphi_k})^\intercal\widetilde{\Sigma}_{\varphi_k}^{-1} -\widetilde{\Sigma}_{\varphi_k}^{-1}\right)\log\left[1 - \Phi\left(\boldsymbol{x}_i^\intercal\boldsymbol{\varphi}_k\right)\right]\right\} \notag\\
    &- \frac{\nu_k}{2\omega_k}I_P + \frac{1}{2}\widetilde{\Sigma}_{\varphi_k}^{-1}. \label{eqn:variational_sigma}
\end{align}
\item $\hat{q}(\boldsymbol{\varphi}_k^0) = \text{Normal}\left(\Tilde{\boldsymbol{\theta}}_{\varphi^0_k}, \widetilde{\Sigma}_{\varphi^0_k}\right)$ for $k=1,\dots, M_w$, where $\Tilde{\boldsymbol{\theta}}_{\varphi^0_k}$ and $ \widetilde{\Sigma}_{\varphi^0_k}$ are defined by:
\begin{align}
	\Tilde{\boldsymbol{\theta}}_{\varphi_k^0} &= \frac{\Tilde{\nu}_k \Tilde{\boldsymbol{\theta}}_{\varphi_k} + \Tilde{\omega}_k\boldsymbol{\mu}}{\Tilde{\nu}_k + \Tilde{\omega}_k}, \label{eqn:variational_theta0} 
    \\
    \widetilde{\Sigma}_{\varphi_k^0} &= \frac{\Tilde{\omega}_k}{\Tilde{\nu}_k + \Tilde{\omega}_k}I_P \label{eqn:variational_sigma0}.
\end{align}
\item $\hat{q}(\sigma_k^2) = \text{Inverse\text{-}Gamma}(\Tilde{\nu}_k, \Tilde{\omega}_k)$ for $k=1,\dots,M_w$, where $\Tilde{\nu}_k$ and $\Tilde{\omega}_k$ are defined by:
\begin{align}
	\Tilde{\nu}_k &= \nu_0 + \frac{P}{2} \label{eqn:variational_nu}, \\
    \Tilde{\omega}_k &= \omega_0 + \frac{1}{2}\left(\Tilde{\boldsymbol{\theta}}_{\varphi_k} - \Tilde{\boldsymbol{\theta}}_{\varphi_k^0}\right)^\intercal\left(\Tilde{\boldsymbol{\theta}}_{\varphi_k} - \Tilde{\boldsymbol{\theta}}_{\varphi_k^0}\right) + \frac{1}{2}\text{tr}\left(\widetilde{\Sigma}_{\varphi_k}\right) + \frac{1}{2}\text{tr}\left(\widetilde{\Sigma}_{\varphi_k^0}\right). \label{eqn:variational_omega}
\end{align}
\item $\hat{q}(\gamma_{ks}^{\prime}) = \text{Beta}(\Tilde{\alpha}_{\gamma_{ks}}, \Tilde{\beta}_{\gamma_{ks}})$ for $k = 1, \dots, M_w$ and $s=1,\dots,M_z$, where $\Tilde{\alpha}_{\gamma_{ks}}$ and $\Tilde{\beta}_{\gamma_{ks}}$ are defined by:
\begin{align}
	\Tilde{\alpha}_{\gamma_{ks}} &= 1 + \sum_{\ell=1}^L \sum_{i=1}^N \Tilde{\boldsymbol{\phi}}_{z_{\ell i},s} \Tilde{\boldsymbol{\phi}}_{w_i,k}, \label{eqn:variational_alpha_gamma}\\
	\Tilde{\beta}_{\gamma_{ks}} &= \eta_0 + \sum_{r=s+1}^\infty \sum_{\ell=1}^L \sum_{i=1}^N \Tilde{\boldsymbol{\phi}}_{z_{\ell i},r} \Tilde{\boldsymbol{\phi}}_{w_i,k}. \label{eqn:variational_beta_gamma}
\end{align}
\item $\hat{q}(\rho_{km}) = \text{Beta}(\Tilde{\alpha}_{\rho_{km}}, \Tilde{\beta}_{\rho_{km}})$ for $k,m=1,\dots,M_z$, where $\Tilde{\alpha}_{\rho_{km}}$ and $\Tilde{\beta}_{\rho_{km}}$ are defined by:
\begin{align}
	\Tilde{\alpha}_{\rho_{km}} &= \alpha_0 + \sum_{\ell = 1}^L \sum_{i,j\in\mathcal{V}, i\neq j} A_{\ell i j}\Tilde{\boldsymbol{\phi}}_{z_{\ell i},k} \Tilde{\boldsymbol{\phi}}_{z_{\ell j},m}, \label{eqn:variational_alpha_rho}\\
	\Tilde{\beta}_{\rho_{km}} &= \beta_0 + \sum_{\ell = 1}^L \sum_{i,j\in\mathcal{V}, i\neq j} (1 - A_{\ell i j}) \Tilde{\boldsymbol{\phi}}_{z_{\ell i},k} \Tilde{\boldsymbol{\phi}}_{z_{\ell j},m}. \label{eqn:variational_beta_rho}
\end{align}
\end{enumerate}

The gradients used to update $\Tilde{\boldsymbol{\theta}}_{\varphi_k}$ and $\widetilde{\Sigma}_{\varphi_k}$ involve computing the inverse of $\widetilde{\Sigma}_{\varphi_k}$, which requires that $\widetilde{\Sigma}_{\varphi_k}$ be invertible. To ensure this, we reparameterise using the log-Cholesky parameterisation. The Cholesky decomposition of $\widetilde{\Sigma}_{\varphi_k}$ gives a lower-diagonal matrix $L_k$ such that $\widetilde{\Sigma}_k = L_kL_k^\intercal$. As $\widetilde{\Sigma}_{\varphi_k}$ is real, $L_k$ is constrained to have positive diagonal entries. We thus introduce an unconstrained matrix $B_k$, corresponding to the log-Cholesky decomposition of $\widetilde{\Sigma}_{\varphi_k}$,   defined as
\begin{equation}
M_{B,ij} = 
\begin{cases}
    \log \left(L_{k, ij}\right), &  i = j, \\
    L_{k, ij}, & i > j,\\
    0 & i < j.
\end{cases}
\end{equation}
This is equivalent to writing
\begin{equation}
B_k = \lfloor L_k\rfloor + \log\left\{\text{diag}(L_k)\right\}, \quad \text{and} \quad 
L_k = \lfloor B_k \rfloor + \exp\left\{\text{diag}(B_k)\right\},
\label{eq:lk}
\end{equation}
where we write $\lfloor \cdot \rfloor$ to denote the strictly lower-diagonal part of a matrix. The log-Cholesky is preferred over a standard Cholesky parametrisation due to its superior numerical stability. Derivatives of the ELBO with respect to $B_k$ then become
\begin{align}
    \pdv{\mathcal{L}}{B_{k, ij}} = \sum_{r,s = 1}^P \pdv{\widetilde{\Sigma}_{{\varphi_k}, rs}}{B_{k, 
            ij}}\pdv{\mathcal{L}}{\widetilde{\Sigma}_{{\varphi_k}, rs}} = \text{tr}\left\{\left(\pdv{\mathcal{L}}{\widetilde{\Sigma}_{\varphi_k}}\right)^\intercal\pdv{\widetilde{\Sigma}_{\varphi_k}}{B_{k, ij}}\right\}.
\end{align}
From \eqref{eq:lk}, we can write
$
\widetilde{\Sigma}_{\varphi_k} = \left(\lfloor B_k \rfloor + \exp\left\{\text{diag}(B_k)\right\}\right)\left(\lfloor B_k \rfloor + \exp\left\{\text{diag}(B_k)\right\}\right)^\intercal
$, 
which gives:
\begin{align}
    \pdv{\widetilde{\Sigma}_{\varphi_k}}{B_{k, ij}} = 
    \begin{cases}
        \exp\{B_{k, ii}\}\left(E_{ii} L_k^\intercal + L_k E_{ii}\right), & i = j, \\
        E_{ij}L_k^\intercal + L_kE_{ji}, & i > j, \\
        0, & i < j, 
    \end{cases}
\end{align}
where we define $E_{ij}$ to be the matrix with element $(r,s)$ equal to $\mathbb{I}_{(i,j)}\{(r,s)\}$.

In CAVI, parameter updates are carried out sequentially conditional upon all others. For consistency, we thus do the gradient steps sequentially, by iterating over $k \in [M_w]$, first updating $\Tilde{\boldsymbol{\theta}}_{\varphi_k}$ and then $\widetilde{\Sigma}_{\varphi_k}$, as opposed to jointly updating $\Tilde{\boldsymbol{\theta}}_{\varphi_k}$ and $\widetilde{\Sigma}_{\varphi_k}$. The full procedure is outlined in Algorithm \ref{alg:full_update_algorithm} in Appendix \ref{app:full_algorithm}.

It should be noted that the $\boldsymbol{\varphi}_k$ will not be recoverable. The map $\alpha \mapsto \Phi(\alpha)$ is numerically non-invertible in the sense that it is not in practice a one-to-one function. For example, we have $\Phi(5)$ and $\Phi(6)$ agree to six decimal places. As the inference only ever processes the image of $\boldsymbol{x}_i^\top\boldsymbol{\varphi}_k$ under this mapping, we cannot hope to recover $\boldsymbol{\varphi}_k$. Furthermore, it is implicitly assumed that the features exhibit a group structure. In the case that the number of these groups is less than $P$, the design matrix $X$ will have an effective rank that is non-maximal. The result of this is that the map $\boldsymbol{\varphi}_k \mapsto X\boldsymbol{\varphi}_k$ has non-trivial nullspace, and so we cannot hope to recover $\boldsymbol{\varphi}_k$ uniquely. Furthermore, although the stick–breaking process imposes an order on the breaks, it does not fix a meaningful scale on the parameters as they are re-normalised in the inference. This, however, is not problematic for the inference as the primary object is the group structure of the nodes. In Section \ref{sec:sims}, we observe that the covariates do aid inference, despite being unable to recover $\boldsymbol{\varphi}_k$.



\subsection{Initialisation}
\label{subsec:initialisation}

\cite{mukherjee2018} demonstrate that variational estimates of the SBM fall into the neighbourhood of a local stationary point with high probability, motivating the need for informed initialisation of the algorithmic parameters.
A common preprocessing step for graph adjacency matrices is spectral embedding, wherein the nodes are embedded into a low-dimensional space via eigendecompositions or singular value decompositions \citep{luo2003}. We use the singular value decomposition of the adjacency matrix to initialise $\{\Tilde{\boldsymbol{\phi}}_{z_{\ell i}}\}$. A truncation in the variational approximation to $q(z)$ is made at some threshold $M_z \in \mathbb{N}$, as per Section \ref{subsec:variational_approxs}. We compute the left singular vectors of $A$ and use HDBSCAN \citep{campello_2013} to cluster the nodes into layer-level groups. HDBSCAN is a non-parametric, density-based clustering algorithm that does not require prior specification of the number of clusters. If the number of clusters found by the algorithm exceeds $M_z$, the minimum number of nodes per cluster is increased and the algorithm re-run until the number of clusters is exactly $M_z$. It is also able to detect outliers, which we assign to the cluster that has the smallest Euclidean distance between its centre and the outlier. After assignment in each layer, we use an implementation of a modified Jonker-Volgenant algorithm with no initialisation \citep{crouse2016} to re-enumerate the learned labels so that they are consistent with the enumeration of layer 1. The choice of layer 1 as reference is arbitrary. An analogous process is run to obtain initial values for $\{\Tilde{\boldsymbol{\phi}}_{w_i}\}$.

To initialise $\Tilde{\alpha}_{\rho_{km}}$ and $\Tilde{\beta}_{\rho_{km}}$, we first set $\Tilde{\beta}_{\rho_{km}} = 1$ for all $k,m \in \{1,\dots,M_z\}$. Then for each layer $\ell \in \{1,\dots,L\}$ of the network, we select all nodes $i,j \in \mathcal{V}$ such that, after the above initialisation, $z_{\ell i} = k$ and $z_{\ell j} = m$. We then compute the average of $\{A_{\ell ij} : z_{\ell i} = k, z_{\ell j} = m\}$ and set this equal to $\Tilde{\alpha}_{\rho_{km}} / (\Tilde{\alpha}_{\rho_{km}} + \Tilde{\beta}_{\rho_{km}})$, the posterior mean for $\rho_{km}$ under the variational approximation. This is inverted to obtain $\Tilde{\alpha}_{km}$.

The parameters $\Tilde{\alpha}_{\gamma_{ks}}$ and $\Tilde{\beta}_{\gamma_{ks}}$ are initialised by computing the proportion of nodes $i\in\mathcal{V}$ that have $w_i = k$ and $z_{\ell i} = s$ . This value is set equal to $\Tilde{\alpha}_{\gamma_{ks}} / (\Tilde{\alpha}_{\gamma_{ks}} + \Tilde{\beta}_{\gamma_{ks}})$ and inverted to solve for $\Tilde{\alpha}_{\gamma_{ks}}$, where we again first set $\Tilde{\beta}_{\gamma_{ks}} = 1$.



\section{Simulation studies}
\label{sec:sims}

We evaluate the performance of the inference procedure proposed in Section \ref{sec:inference} using simulated data. In Section \ref{subsec:sim1}, we examine recovery of the global and layer-level groups with different truncation parameters. Section \ref{subsec:sim2}, considers a network with well-separated layer-level behaviour but with increasingly uninformative nodal covariates. Section \ref{subsec:sim3} considers networks with informative features, but decreasing separation between each global group's behaviour across layers. Next, Section \ref{subsec:sim4} examines group recovery with varying network size and varying numbers of layers, in particular considering a setting in which our connectivity matrix has weak separation. We evaluate using the normalised mutual information \citep[NMI;][]{strehl_2002} score, which ranges from 0 to 1, with 0 corresponding to random assignment, and 1 perfect agreement. This metric accounts for labelling differences between simulations. To align the inferred groups with the true simulated groups, we use an implementation of a modified Jonker-Volgenant algorithm with no initialisation \citep{crouse2016}, using the \texttt{linear\_sum\_assignment} method from \texttt{scikit-learn} to re-enumerate the learned labels so that they are consistent with those of the simulation. The number of components are selected by assigning each node to the group corresponding to the index of its posterior probability vector with maximum value.

\subsection{Inference of the number of latent groups}
\label{subsec:sim1}

In this simulation, we demonstrate that our inference procedure can correctly recover the number of global and layer-level groups when the truncations $M_w$ and $M_z$ are larger than the true simulation values. We consider $N=250$ nodes, split in the ratio 3:2 between two global groups. We set the probability vectors over three layer-level groups to be:
\begin{equation}
    \boldsymbol{\gamma}_1 = (0.8, 0.1, 0.1), \quad \boldsymbol{\gamma}_2 = (0, 0.5, 0.5),
\end{equation}
and sample their features from multivariate normals with identity covariance matrices and mean vectors
\begin{equation}
    \boldsymbol{\mu}_1 = (3/2,3/2,3/2), \quad \boldsymbol{\mu}_2 = (-3/2,-3/2,-3/2).
\end{equation}
We set the probability matrix $\rho$ to be
\begin{equation}
    \rho = 
    \begin{pmatrix}
        0.8 & 0.5 & 0.2 \\
        0.4 & 0.7 & 0.05 \\
        0.2 & 0.01 & 0.6
    \end{pmatrix},
\end{equation}
and run the CAVI inference procedure for 10 iterations, which was found to be sufficient for convergence in the sense that relative changes in consecutive ELBO evaluations fell below a small threshold $\epsilon$. The fast convergence time is attributed to the informed initialisation scheme we implement. We run the procedure 50 times with $M_w = 2$ and $M_z=3$, and again with $M_w=M_z=5$. In Table \ref{tab:nmi_sim2} we see that the NMI for the global groups is very high for both truncation parameter pairs, with a median value of 1. A corresponding table for the layer level groups is not included as the median was found to be 1.0 with a standard deviation of 0, indicating perfect recovery in all simulations.

\begin{table}[htbp]
  \centering
    \begin{tabular}{|c|c|c|c|c|}
      \hline
      $\boldsymbol{(M_z, M_w)}$ & \textbf{Median} & \textbf{Std.} & \textbf{2.5\%} & \textbf{97.5\%} \\
      \hline
      (2,3) & 1.0 & 0.011 & 0.966 & 1.0 \\
      (5,5) & 1.0 & 0.018 & 0.952 & 1.0 \\
      \hline
    \end{tabular}
  \caption{Summary of the NMI scores for the global groups under the settings of Section \ref{subsec:sim1}.}
  \label{tab:nmi_sim1}
\end{table}



\subsection{Decreasing feature separation}
\label{subsec:sim2}

We examine the setting where the distributions over the layer-level groups of each global group are distinct from one another, but the features of the global groups have decreasing separation. We consider $N=500$ nodes split over 3 global groups in the ratio 2:2:1. We set the probability vectors governing the distributions for the global groups over the layer-level groups to be
\begin{equation}
    \boldsymbol{\gamma}_1 = (1, 0, 0), \quad \boldsymbol{\gamma}_2 = (0,1,0), \quad \boldsymbol{\gamma}_3 = (0,0,1).
\end{equation}
The features are sampled from multivariate normal distributions with identity covariance matrix, and mean vectors of the form
\begin{equation}
    \boldsymbol{\mu}_1 = (\alpha, \alpha, \alpha), \quad \boldsymbol{\mu}_2 = (0,0,0), \quad \boldsymbol{\mu}_3 = (-\alpha,-\alpha,-\alpha),
\end{equation}
where $\boldsymbol{\mu}_k$ corresponds to global group $k$. We allow $\alpha$ to vary over the set $\{5/2, 2, 3/2, 1, 1/2, 0\}$, corresponding to decreasing separation with decreasing $\alpha$. The connection probability matrix $\rho$ we set to be as in the simulation of Section \ref{subsec:sim1}. 

In Table \ref{tab:nmi_sim2} we see that the NMI for the global groups is high across all the simulations, with a median value of 1. Even in the case when the the nodal covariates are uninformative, the global group recovery is good. Again, a corresponding table for the layer level groups is not included as the median was found to be 1.0 with a standard deviation of 0, indicating perfect recovery in all simulations. This simulation study illustrates the trade-off between the signal from the nodal covariates and the network structure. In particular, it demonstrates that the network structure is the primary driver in the recovery of the global groups, and that the covariates serve to provide additional signal that aids the inference.


\begin{table}[htbp]
  \centering
    \begin{tabular}{|c|c|c|c|c|}
      \hline
      $\boldsymbol{\alpha}$ & \textbf{Median} & \textbf{Std.} & \textbf{2.5\%} & \textbf{97.5\%} \\
      \hline
      $5/2$ & 1.0 & 0.148 & 0.643 & 1.0 \\
      2 & 1.0 & 0.176 & 0.643 & 1.0 \\
      $3/2$ & 1.0 & 0.179 & 0.643 & 1.0 \\
      1 & 1.0 & 0.139 & 0.643 & 1.0 \\
      $1/2$ & 1.0 & 0.100 & 0.650 & 1.0 \\
      0 & 1.0 & 0.103 & 0.674 & 1.0 \\
      \hline
    \end{tabular}
  \caption{Summary of the NMI scores for the global groups with decreasing $\alpha$ under the simulation setting of Section \ref{subsec:sim2}.}
  \label{tab:nmi_sim2}
\end{table}

\subsection{Increasing similarity of layer behaviour}
\label{subsec:sim3}

\begin{figure}[t]
\centering\includegraphics[width=\textwidth]{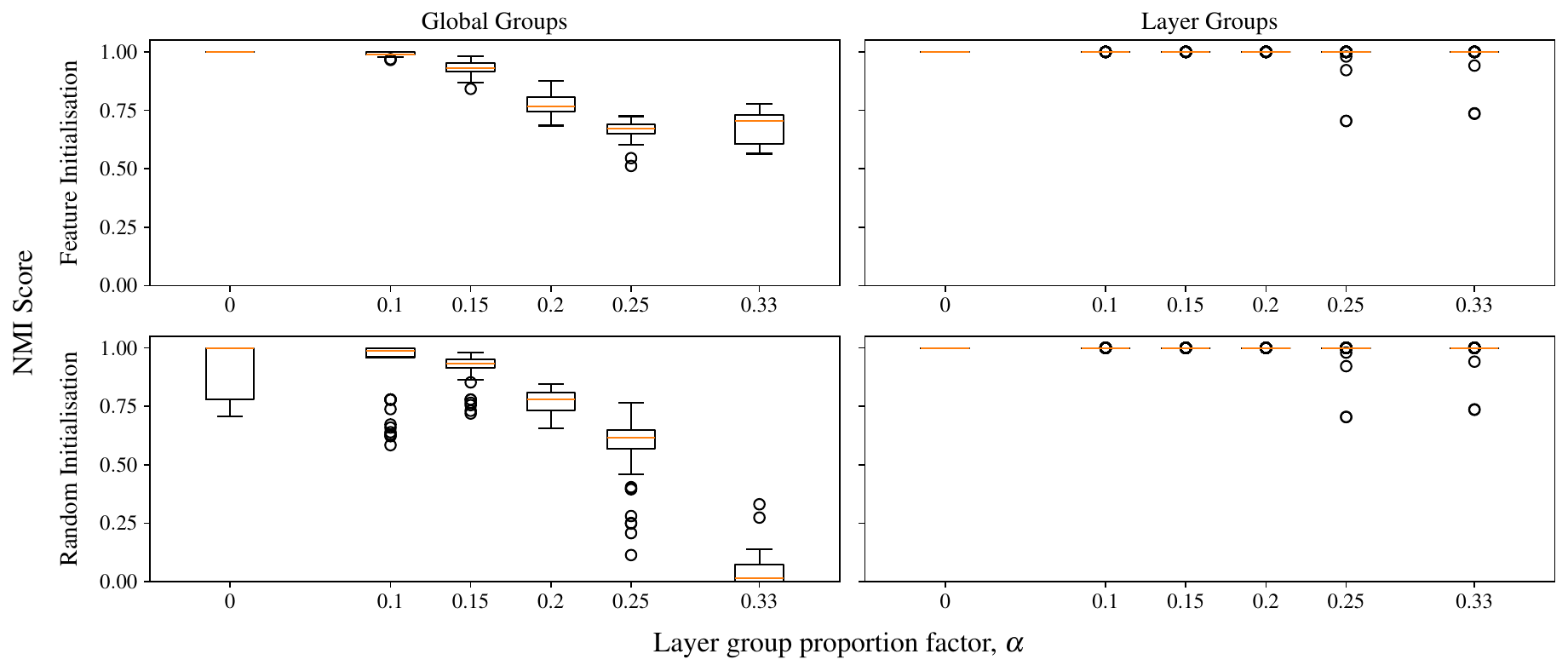}
    \caption{Boxplots of the NMI score with increasing $\alpha$ for the global and layer groups under the settings of Section \ref{subsec:sim3}.}
    \label{fig:NMI_sim_3}
\end{figure}

We consider a setting in which the distributions for the features are distinct between groups, and examine recovery as we increase the similarity between their layer-level behaviour. Specifically, we sample the features from multivariate normal distributions, each with identity covariance matrix, but with the means
\begin{equation}
    \boldsymbol{\mu}_1 = (5,5,5), \quad \boldsymbol{\mu}_2 = (0,0,0), \quad \boldsymbol{\mu}_3 = (-5,-5,-5),
\end{equation}
where $\boldsymbol{\mu}_k$ corresponds to global group $k$. We then set the distributions over the layer-level groups to be:
\begin{equation}
    \boldsymbol{\gamma}_1 = (1 - 2\alpha, \alpha, \alpha), \quad \boldsymbol{\gamma}_2 = (\alpha,1-2\alpha,\alpha), \quad \boldsymbol{\gamma}_3 = (\alpha,\alpha,1-2\alpha).
\end{equation}
We again set $\rho$ as in the simulation of Section \ref{subsec:sim2}. We consider $N=500$ nodes split over 3 global groups in the ratio 2:2:1 and allow for 25 iterations of the full CAVI scheme, with a maximum of 30 gradient ascent steps per variable per iteration. We run the simulation study twice: once when $\{\Tilde{\boldsymbol{\phi}}_{w_i}\}_{i\in \mathcal{V}}$ is initialised according to the method in Section \ref{subsec:initialisation}, which we call informed initialisation, and once when each component is $1/M_w$, corresponding to an uninformed initialisation. As was discussed in Section \ref{sec:model}, due to inherent identifiability issues in the Dirichlet process and the impact of starting values on the CAVI algorithm, we cannot expect different initialisations to converge to the same values on all parameters. However, we demonstrate that this doesn't impact our results. Furthermore, in Figure \ref{fig:NMI_sim_3}, it is seen that the informed initialisation leads to NMI values with a higher median and smaller standard deviation in each simulation setting when compared to random initialisation.

This simulation study further examines the balance between the nodal covariates and the network structure. We expect the network structure to dominate the covariates in the inference. In Figure \ref{fig:NMI_sim_3} we see that the layer groups are recovered well in the sense of NMI in all simulations. The NMI is high for the global groups for small values of $\alpha$ and decreases as the similarity between the layers increases. We see very similar recovery for the informed and uninformed initialisations, except for when $\alpha=0.33$, which corresponds to the behaviour of each global group being identical across layers. This simulation study reaffirms the conclusion of Section \ref{subsec:sim2} in that the inference on the global groups is driven primarily by the network structure, with the nodal covariates serving as an additional guide.

\subsection{Effect of the number of layers and network size}
\label{subsec:sim4}

\begin{figure}[t]
\centering\includegraphics[width=\textwidth]{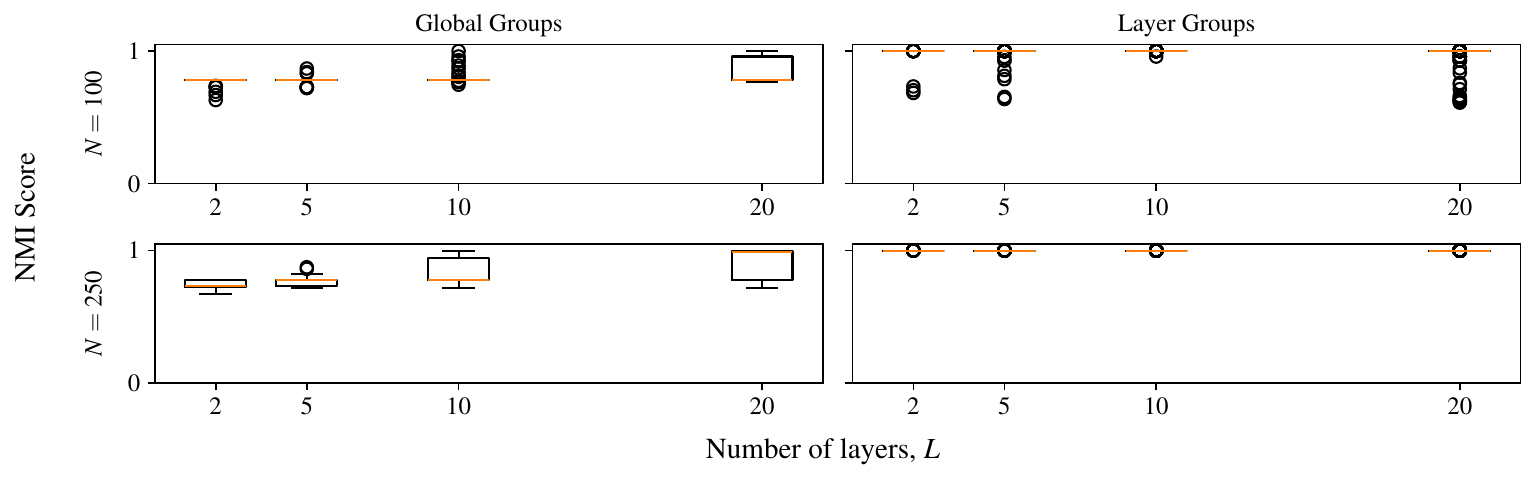}
    \caption{Boxplots of the NMI score with varying number of nodes, $N$, and varying numbe of samples, $M$, for the global and layer groups under the settings of Section \ref{subsec:sim4}.}
    \label{fig:NMI_sim_4}
\end{figure}

This simulation examines the effect of the number of layers and number of nodes on parmeter recovery. We select $\rho$ as in Section \ref{subsec:sim1} and feature and global group distributions as in Section \ref{subsec:sim3} with $\alpha = 0.15$. This corresponds to good feature separation but global groups that exhibit reasonably similar behaviour. We consider $N \in \{100,250\}$ and $L \in \{2,5,10,20\}$. We run the CAVI inference procedure for 10 steps with $M_w=M_z=3$. Boxplots of the resulting NMI output is show in Figure \ref{fig:NMI_sim_4}. We see increasing performance with increasing $L$ and for both values of $N$, as would be expected. The layer groups are perfectly recovered for $N=250$ for all values of $L$. For $L=2,5,10$, the global group recovery is comparable between $N=100$ and $N=250$, but for $L=20$, the larger network exhibits notably better recovery. 

\section{Application to the FAO trade network}
\label{sec:real_data}

The proposed model and inference procedure were deployed on the Food and Agriculture Organization of the United Nations (FAO) trade network worldwide food import and export network from 2010. \cite{domenico2015} collected data from the FAO web API and transformed it into a multiplex network with countries as nodes and layers with their respective edges representing import and export relationships between countries for a specific food. The resulting graph is directed, corresponding to import and export. The raw network consists of $N=214$ nodes (countries) and $L=314$ layers (food types).

This dataset was also analysed in \cite{amini2024}, and we follow the same pre-processing steps. The network provides weighted edges, but we transform to a binary adjacency matrix by converting any non-zero weight to 1. We select the 20 layers with the highest edge count and then remove any nodes that have fewer than 20 edges in the aggregated adjacency matrix, corresponding to the sum of the adjacency matrices of each of the 20 most dense layers. The resulting multiplex network has $N=177$ nodes and $L=20$ layers.

The network does not contain nodal covariates, and so we source features from the World Bank \citep{worldbank2023gdp}. We run the inference twice, once using the GDP from 2009 of each country and its rural land area in square kilometers, and another using these two features, but additionally urban land area and coast-line length. We use a log-transformation before normalising to have zero mean and unit variance. We refer to these settings as setting A and setting B, respectively. We focus primarily on setting A, but will consider the results of setting B for comparison. We also include an intercept term for each node. For some countries, such as Macao and the Netherlands Antilles (which was dissolved in 2010 into separate Caribbean territories), we were able to get only one feature. We sorted each feature, found the rank of the feature we could obtain and used that ranking to impute the missing value of the other feature. One country is named \texttt{Unspecified}, and for this we impute the values of the covariates using the global median. The algorithm of Section \ref{sec:inference} was run with varying truncation parameters, and we found that the number of layer-level groups consistently collapsed onto 6. We thus set $M_w = 12$ and $M_z=6$ in the inference and allow for 25 iterations of the CAVI procedure with a maximum of 30 gradient ascent steps per parameter. The algorithm detected 11 global groups and 6 layer-level groups (the same numbers were found with higher truncation values).

In Figure \ref{fig:global_layer_level_proportions} we see the distribution of layer-level groups within each global group for setting A. This corresponds to an estimate of $\{\boldsymbol{\gamma}_k\}_{k=1}^{11}$ in the model. We see that the behaviour of each global group is distinct from one another, with different global groups having different distributions across the layer-level groups. 



    
    
    

\if1\jcn
{
\begin{figure}[htbp]
  \centering
  \subfigure[Rural land area and GDP.]{
    \includegraphics[width=\linewidth]{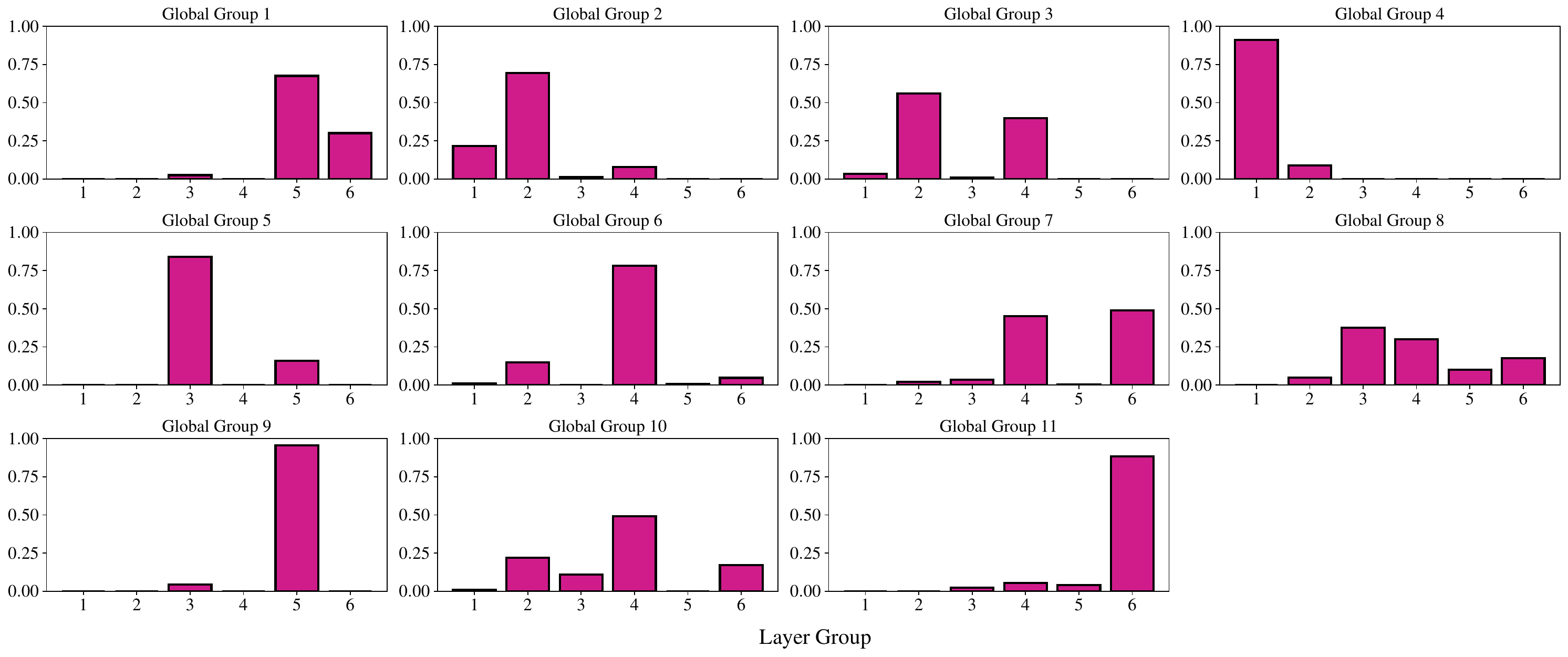}
    \label{fig:global_layer_level_proportions}
  }
  \par\addvspace{0.8em} 
  \subfigure[Rural land area, GDP, urban land area and coast-line length.]{
    \includegraphics[width=\linewidth]{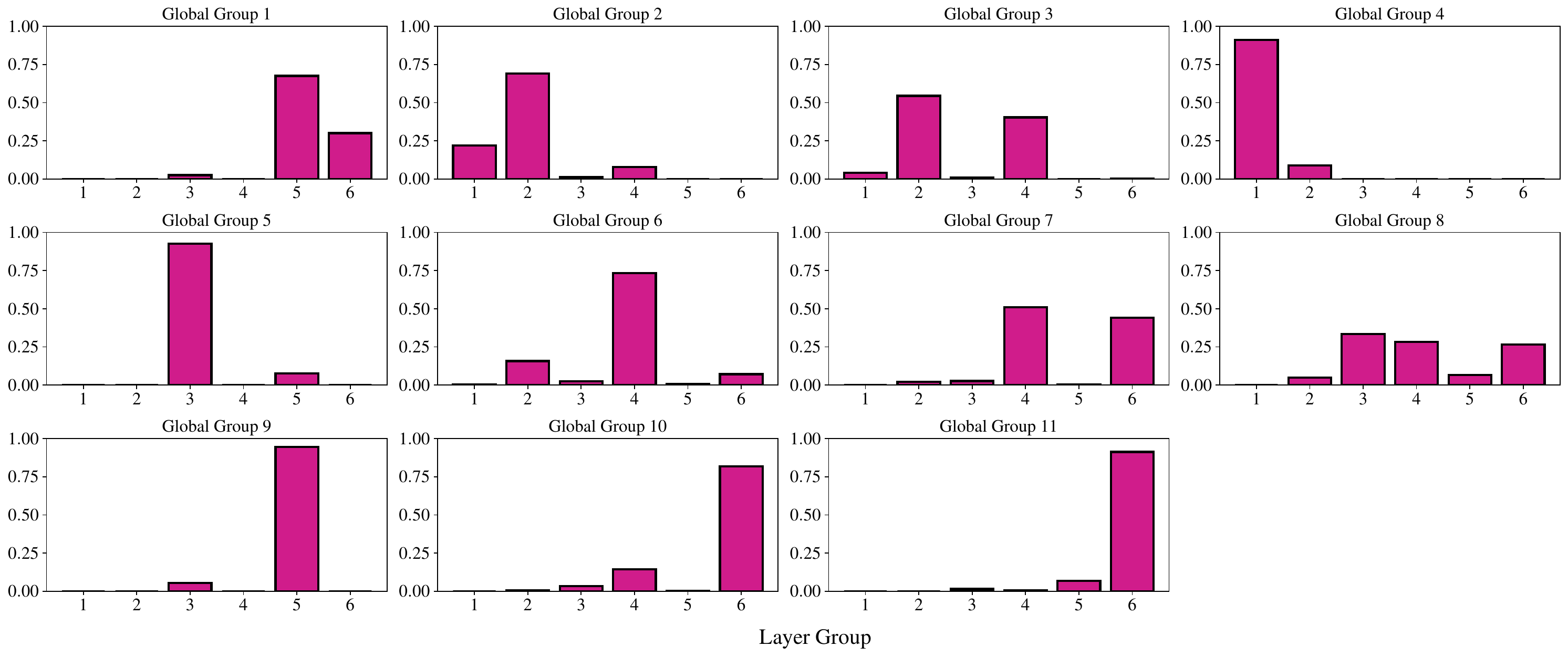}
    \label{fig:global_layer_level_proportions_add}
  }
  \caption{Distribution of layer-level groups in each global group using different features.}
  \label{fig:layer_group_dist}
\end{figure}
}\else{
\begin{figure}[t]
  \centering
  \begin{subfigure}{\linewidth}
    \centering
    \includegraphics[width=\linewidth]{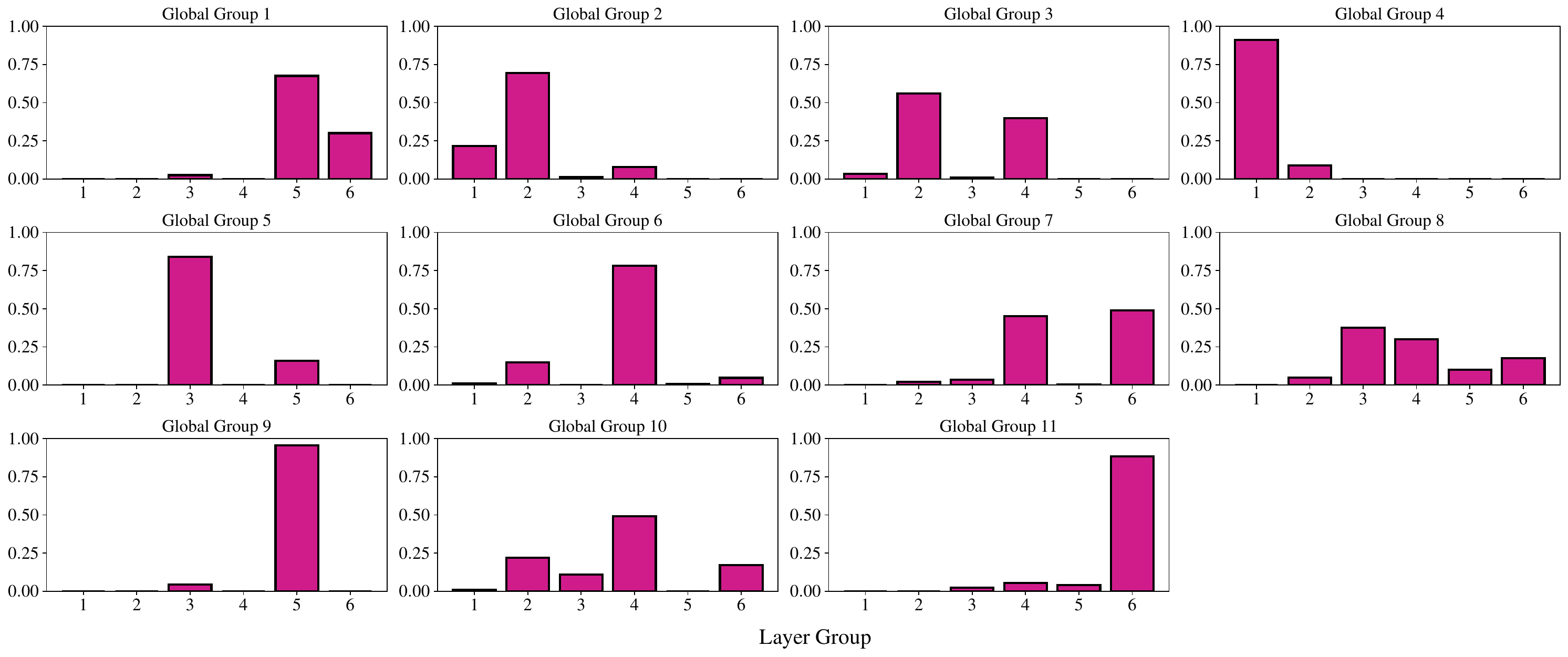}
    \caption{Rural land area and GDP.}
    \label{fig:global_layer_level_proportions}
  \end{subfigure}

  \vspace{0.8em} 

  \begin{subfigure}{\linewidth}
    \centering
    \includegraphics[width=\linewidth]{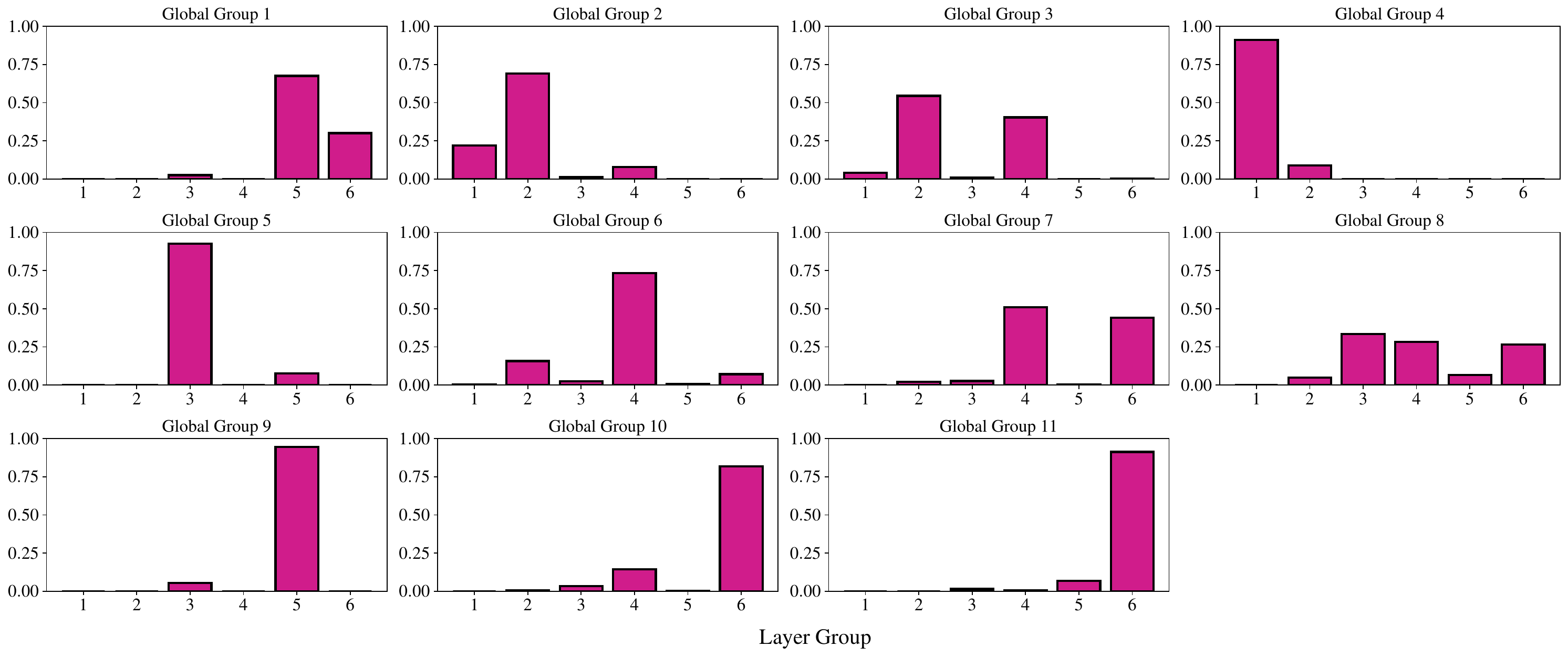}
    \caption{Rural land area, GDP, urban land area and coast-line length.}
    \label{fig:global_layer_level_proportions_add}
  \end{subfigure}

  \caption{Distribution of layer-level groups in each global group using different features.}
  \label{fig:layer_group_dist}
\end{figure}
}\fi

\begin{figure}[t]
\centering\includegraphics[width=\textwidth]{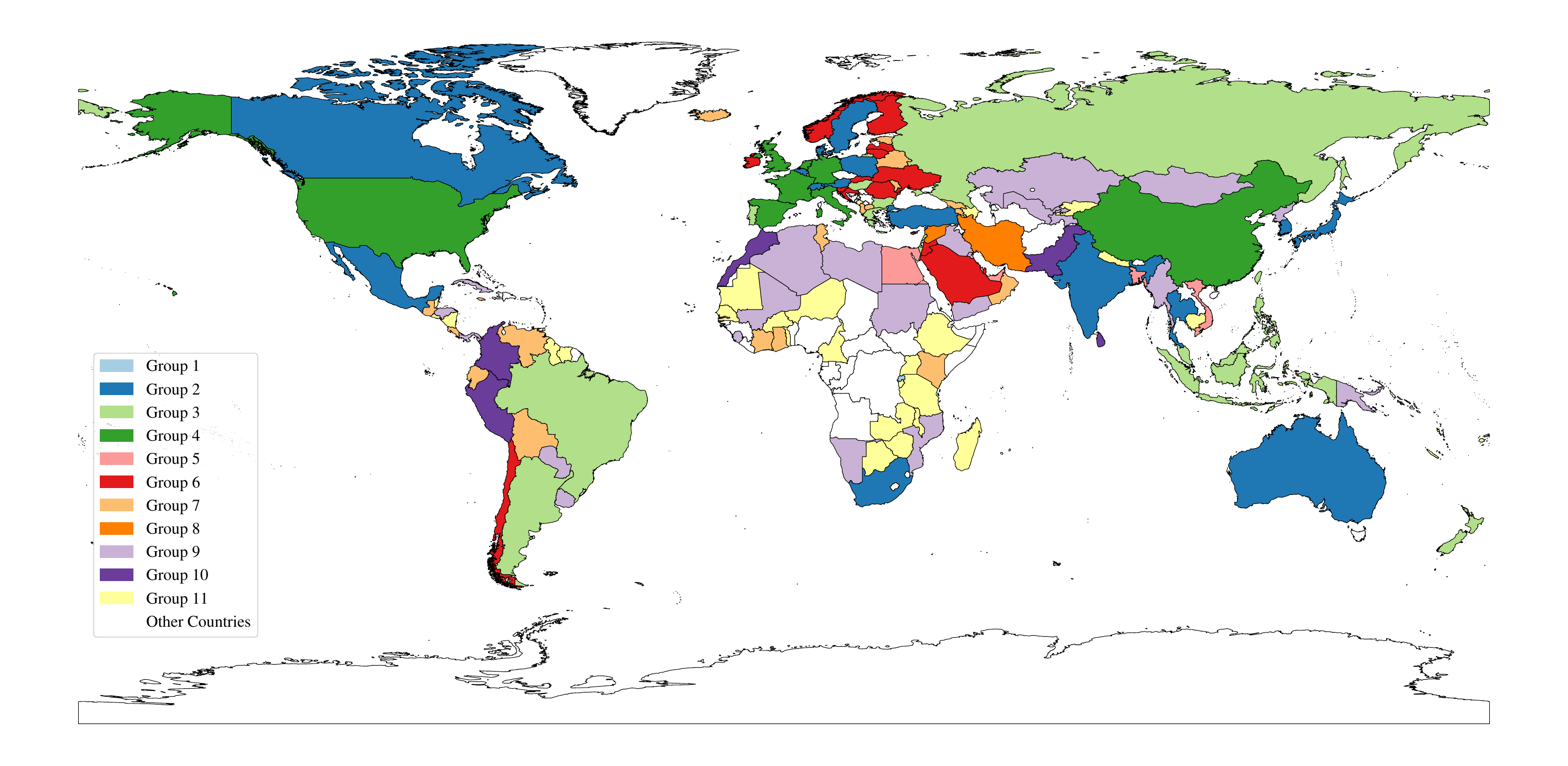}
    \caption{Map of the world with countries present in the processed network coloured according to their inferred global group.}
    \label{fig:world_map_FAO}
\end{figure}

In Figure \ref{fig:world_map_FAO} is a world map with countries that were present in the network coloured by their inferred global group for setting A. Group 4 (dark green) consists of France, China (main land), the United States of America, the United Kingdom, Italy, Spain, the Netherlands and Germany. This is an intuitive community as they represent some of the largest global economies and trading powers at the time. Group 3 (light green) consists of many emerging economies and regional powers (Brazil, Russia, Argentina) plus some smaller EU economies (Portugal, Greece) and Asian trading centers (Hong Kong, Malaysia). Group 2 seems to sit between these two groups, consisting of a mixture of advanced economies (Japan, Switzerland, Denmark, Sweden, Canada, South Korea) and emerging economies with major agricultural output (India, Turkey, Mexico, South Africa). A particularly interesting community is the group 8 (dark orange) which is made-up of only Syria and Iran. The dataset in question is from the year before the Syrian Revolution of 2011. Iran was one of the largest financial supporters of then President Bashar al-Assad and so it is interesting to see that our inference procedure has clustered these countries together. 

We see that some of the global groups consist of countries that are geographically close. Group 6 (dark red) contains a sizeable cluster of Eastern European and Baltic states (Romania, Lithuania, Slovakia, Slovenia, Serbia, Latvia). Similarly, group 9 (light purple) contains a subgroup of Central Asian countries (Kazakhstan,  Uzbekistan, Tajikistan) and another of North African/Middle Eastern countries (Algeria, Libya, Iraq, Kuwait, Yemen). However, the communities clearly do not only correspond to countries that are close to one another.  

\begin{figure}[t]
\centering\includegraphics[width=\textwidth]{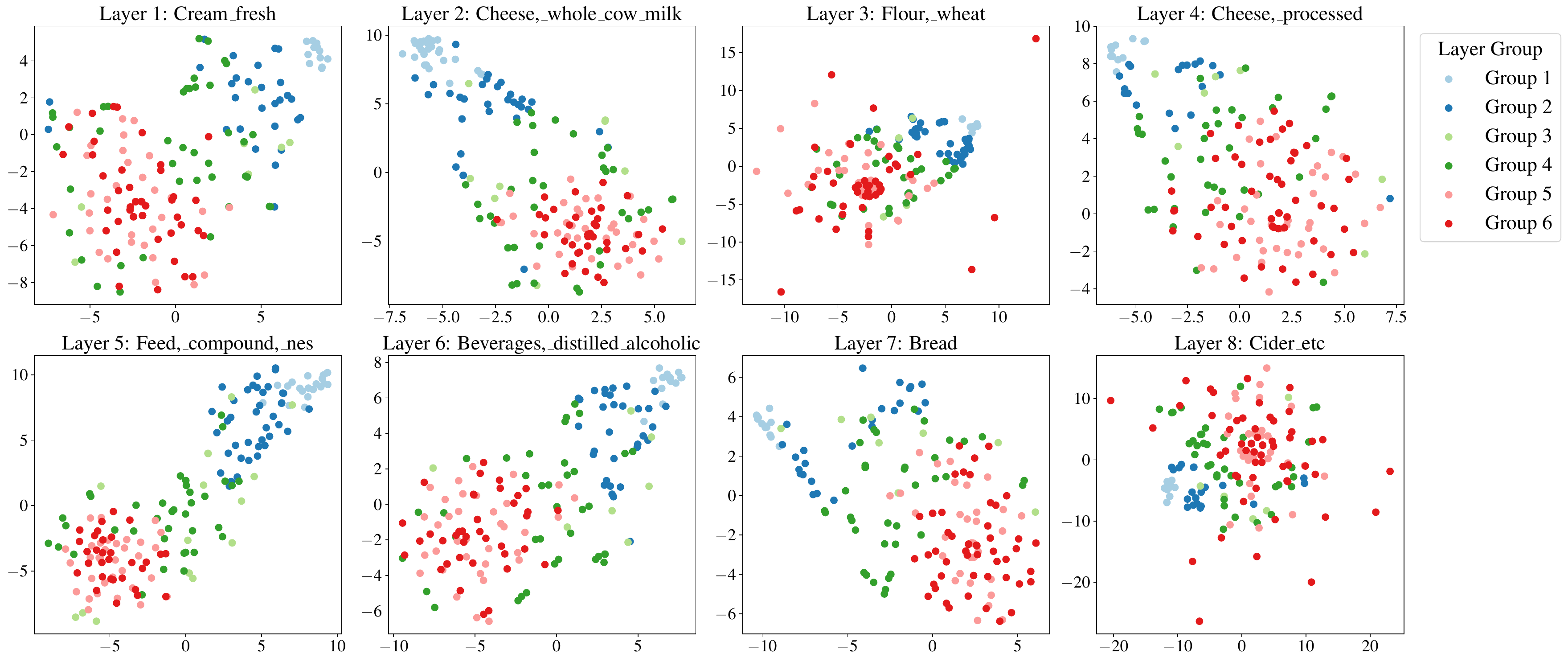}
    \caption{t-SNE plots in two dimensions for layers 1 to 8 with nodes coloured according to their layer-level group.}
    \label{fig:tsne_plot}
\end{figure}

We then use t-Distributed Stochastic Neighbor Embedding (t-SNE), a popular non-linear dimension reduction tool, to visualise our layer-level clustering for setting A. Figure \ref{fig:tsne_plot} shows plots of a two-dimensional t-SNE embedding for layers 1 to 8 of the network using a perplexity of 30, with nodes coloured according to their layer-level group. We see that nodes of the same colour tend to appear near to one another the embedding space, and that this holds in all of the layers of the network. This demonstrates that the clustering has effectively uncovered latent structure in the multiplex network. An analogous plot for all layers of the network is provided in Figure \ref{fig:tsne_all_layer} in Appendix \ref{app:additional_figures}. 

Most of the panels in Figure \ref{fig:tsne_plot} (with the exception of Layer 8) show a conic shape, with the blue groups (layer groups 1 and 2) showing low variance, the reds (layer groups 5 and 6) high variance and the greens (layer groups 3 and 4) sitting between the two. In Figure \ref{fig:global_layer_level_proportions}, we see that global groups 2 and 4 are mostly concentrated on layer groups 1 and 2. These global groups correspond mostly to advanced economies, which explains their low variability. Similarly, global groups 1, 9 and 11 are emerging economies in Africa and are concentrated on layer groups groups 5 and 6, which corresponds to the higher variance of the clusters.

We now examine the output for setting B. After aligning labels, the global-group NMI is 0.89 when compared to setting A, indicating very high agreement between the two fits. Figures~\ref{fig:global_layer_level_proportions} and \ref{fig:global_layer_level_proportions_add} show that the layer group distributions within each global group are nearly unchanged between settings A and B, with the main difference in Group 10. 
    
In total, 29 of the 177 countries (16.4\%) changed global group after alignment, with 17 moving from Group 10 to Group 9. The countries driving this shift are mostly coastal/island states, consistent with the addition of coastline and urbanisation features. Four other countries moved into group 9 (Albania, Bolivia, Jamaica, Venezuela), three of which are coastal/island, again consistent with coastline length drawing them into a coastal cluster. We see stability within the developed-economy blocks. Importantly, the group corresponding to larger, more developed economies is unchanged by the added covariates: United States, United Kingdom, Germany, France, Italy, Spain, China, and the Netherlands all remain in the same global group as before. Likewise, other major economies such as Canada, Australia, Japan, Denmark, Sweden, and Switzerland also retain their original assignments. This indicates that the covariate augmentation refines the periphery (e.g., coastal/island states) while leaving the core, high-income structure intact that is predominantly determined by the network structure. Overall, the resulting change from setting A to setting B suggest that network structure remains the primary driver of the clustering, with the new covariates making small and interpretable refinements.

\section{Conclusion}

We have presented the hierarchical multiplex stochastic blockmodel (HMPSBM), a novel Bayesian model and inference framework for simultaneous layer-level and global clustering of nodes in a multiplex network, assuming SBMs in each layer. We leverage a non-parametric Dirichlet process prior to allow for inference of number of latent groups without prior specification. Our variational inference procedure is fast and scalable. We have tested our inference procedure on both real and simulated data. For the simulated data, we see that our methodology is able to capture the latent structure accurately. In the case of the real data, we uncover interesting latent structure among countries in the FAO trade network. Our global groupings are mostly intuitive within the context of the data.

There are a number of ways in which the model could be extended. In particular, an extension to the case where nodes are not required to be present in every layer would be interesting, wherein the frequency of absence across layers could contribute to the node's global clustering. Furthermore, an extension to the setting where covariates are not fully observed would allow for application of the model in wider settings.

\section*{Acknowledgements}

Joshua Corneck acknowledges funding from the Engineering and Physical Sciences Research Council (EPSRC), grant number EP/S023151/1.
Ed Cohen acknowledges funding from the EPSRC NeST Programme, grant number EP/X002195/1.
Francesco Sanna Passino acknowledges funding from the EPSRC, grant number EP/Y002113/1. 

\ \\
This paper describes objective technical results and analysis. Any subjective views or opinions that might be expressed in the paper do not necessarily represent the views of the U.S. Department of Energy or the United States Government. This work was supported by the Laboratory Directed Research and Development program at Sandia National Laboratories, a multimission laboratory managed and operated by National Technology and Engineering Solutions of Sandia, LLC, a wholly-owned subsidiary of Honeywell International, Inc., for both the U.S. Department of Energy’s National Nuclear Security Administration under contract DE-NA0003525.
\section*{Code}

\textit{Python} code to implement the methodologies proposed in this article and reproduce the results is available in the Github repository {\color{red}\href{https://github.com/joshcorneck/HMPSBM.git}{\texttt{joshcorneck/HMPSBM}}}.


\bibliographystyle{rss}  
\bibliography{references} 

\newpage
\begin{appendices}

\section{Complete conditional distributions} \label{app:complete_conditional}

Adopt the notation $\Theta$ for the full collection of latent variables, and $\Theta_{-\boldsymbol{u}}$ for the vector $\Theta$ with $\boldsymbol{u}$ removed, for example. In this appendix, the forms of the complete conditional distributions are derived.

\subsection{Complete conditional distribution for \texorpdfstring{$z_{\ell i}$}{z}}
\label{app:complete_conditional_zli}

The complete conditional distribution for $z_{\ell i}$ can be written as
\begin{equation}
p(z_{\ell i} \mid \Theta_{-z_{\ell i}}, A, X) \propto p(A \mid z, \rho) \times p(z \mid \boldsymbol{w}, \gamma^{\prime}),
\label{eq:complete_cond_z}
\end{equation}
The term $p(A \mid z, \rho)$ factorises over $\ell$ but not over $i$ and $j$, and we need to retain in the equation only the terms depending on $z_{\ell i}$. Therefore, within equation \eqref{eq:complete_cond_z}, we have: 
\begin{align}
    p(A \mid z, \rho) &\propto \prod_{j\in [N] \setminus \{i\}} \Bigg[ \rho_{z_{\ell i} z_{\ell j}}^{A_{\ell ij}}(1 - \rho_{z_{\ell i}z_{\ell j}})^{1 - A_{\ell i j}}\rho_{z_{\ell j}z_{\ell i}}^{A_{\ell ji}}(1 - \rho_{z_{\ell j}z_{\ell i}})^{1 - A_{\ell ji }}\Bigg],\\
    p(z \mid \boldsymbol{w}, \gamma^{\prime}) &= \prod_{\ell=1}^L \prod_{i=1}^N \prod_{s=1}^\infty (\boldsymbol{\gamma}_{w_i})_s^{\mathbb{I}\{z_{\ell i} = s\}} \propto \prod_{s=1}^\infty (\boldsymbol{\gamma}_{w_i})_s^{\mathbb{I}\{z_{\ell i} = s\}}.
\end{align}
Combining these terms, it follows
\begin{align}
    p(z_{\ell i} \mid \Theta_{-z_{\ell i}}, A, X) \propto& \prod_{k=1}^\infty\prod_{w=1}^\infty (\boldsymbol{\gamma}_w)_z^{\mathbb{I}\{w_i=w, z_{\ell i} = k\}} \times \prod_{j\in[N]\setminus\{i\}} \prod_{m=1}^\infty\bigg[\rho_{km}^{A_{\ell ij}}(1 
    -\rho_{km})^{1 - A_{\ell i j}}\\
    &\times\rho_{mk}^{A_{\ell ji}}(1 - \rho_{mk})^{1 - A_{\ell ji}}\bigg]^{\mathbb{I}\{z_{\ell i} = k, z_{\ell j} = m\}}, \\
    =& \prod_{k=1}^\infty\Bigg\{\prod_{w=1}^\infty (\boldsymbol{\gamma}_w)_z^{\mathbb{I}\{w_i=w\}} \times \prod_{j\in[N]\setminus\{i\}} \prod_{m=1}^\infty\bigg[\rho_{km}^{A_{\ell ij}}(1 
    -\rho_{km})^{1 - A_{\ell i j}}\\
    &\times\rho_{mk}^{A_{\ell ji}}(1 - \rho_{mk})^{1 - A_{\ell ji}}\bigg]^{\mathbb{I}\{z_{\ell j} = m\}}\Bigg\}^{\mathbb{I}\{z_{\ell i} = k\}}
\end{align}
and so the complete conditional of $z_{\ell i}$ is categorical, written
\begin{align}
    z_{\ell i} \mid \Theta_{-z_{\ell i}}, A, X \sim \text{Categorical}(\boldsymbol{\phi}_{z_{\ell i}}),
    \label{eqn:complete_conditional_z}
\end{align}
where
\begin{align}
    (\boldsymbol{\phi}_{z_{\ell i}})_k &\propto \prod_{w=1}^\infty (\boldsymbol{\gamma}_w)_z^{\mathbb{I}\{w_i=w\}} \prod_{j\in[N]\setminus\{i\}} \prod_{m=1}^\infty\bigg[\rho_{km}^{A_{\ell ij}}(1 
    -\rho_{km})^{1 - A_{\ell i j}}\rho_{mk}^{A_{\ell ji}}(1 - \rho_{mk})^{1 - A_{\ell ji}}\bigg]^{\mathbb{I}\{z_{\ell j} = m\}}.
    \label{eqn:phi_z}
\end{align}

\subsection{Conditional distribution for \texorpdfstring{$w_i$}{w}} \label{app:complete_conditional_wi}

The complete conditional distribution for $w_{i}$ takes for the form
\begin{equation}
p(w_i \mid \Theta_{-w_i}, A, X) \propto p(z \mid \boldsymbol{w}, \boldsymbol{\gamma}^{\prime}) \times p(\boldsymbol{w} \mid \boldsymbol{\varphi}, X).
\label{eq:complete_cond_w}
\end{equation}
We look to isolate all terms pertaining to $w_i$ in the two terms in equation \eqref{eq:complete_cond_w}, obtaining: 
\begin{align}
    p(z \mid \boldsymbol{w}, \boldsymbol{\gamma}^{\prime}) &= \prod_{\ell=1}^L \prod_{j=1}^N \prod_{s=1}^\infty (\boldsymbol{\gamma}_{w_j})_k^{\mathbb{I}\{z_{\ell j} = k\}} \propto \prod_{\ell=1}^L \prod_{k=1}^\infty (\boldsymbol{\gamma}_{w_i})_k^{\mathbb{I}\{z_{\ell i} = k\}} = \prod_{\ell = 1}^L \prod_{k,w=1}^\infty (\boldsymbol{\gamma}_{w})_k^{\mathbb{I}\{z_{\ell i} = k, w_i = w\}} \\
    p(\boldsymbol{w} \mid \boldsymbol{\varphi}, X) &= \prod_{j=1}^N \prod_{w=1}^\infty (\boldsymbol{\tau}_j)_w^{\mathbb{I}\{w_j = w\}} \propto \prod_{w=1}^\infty (\boldsymbol{\tau}_i)_w^{\mathbb{I}\{w_i = w\}}. 
\end{align}
Combining these expressions, we obtain the following complete conditional distributions for $w_i$:
\begin{align}
    p(w_i \mid \Theta_{-w_i}, A, X) &\propto \prod_{w=1}^\infty (\boldsymbol{\tau}_i)_w^{\mathbb{I}\{w_i=w\}}\times \prod_{\ell=1}^L \prod_{k=1}^\infty (\boldsymbol{\gamma}_w)_k^{\mathbb{I}\{z_{\ell i} = k,w_i = w\}} \\
    &= \prod_{w=1}^\infty \Bigg\{(\boldsymbol{\tau}_i)_w\times \prod_{\ell=1}^L \prod_{k=1}^\infty (\boldsymbol{\gamma}_w)_k^{\mathbb{I}\{z_{\ell i} = k\}}\Bigg\}^{\mathbb{I}\{w_i=w\}},
\end{align}
from which it follows that the complete conditional for $w_i$ is categorically distributed, written
\begin{align}
    w_i \mid \Theta_{-w_i}, A, X \sim \text{Categorical}\left(\boldsymbol{\phi}_{w_i}\right),
    \label{eqn:complete_conditional_w}
\end{align}
where
\begin{align}
    (\boldsymbol{\phi}_{w_i})_w \propto (\boldsymbol{\tau}_i)_w\times \prod_{\ell=1}^L \prod_{k=1}^\infty (\boldsymbol{\gamma}_w)_k^{\mathbb{I}\{z_{\ell i} = k\}}.
    \label{eqn:phi_w}
\end{align}

\subsection{Complete conditional distribution for \texorpdfstring{$\boldsymbol{\varphi}_k$}{phi}}
\label{app:complete_conditional_phik}

The complete conditional for $\boldsymbol{\varphi}_{k}$ can be written as
\[
p(\boldsymbol{\varphi}_{k} \mid \Theta_{-\boldsymbol{\varphi}_{k}}, A, X) \propto p(\boldsymbol{\varphi} \mid \boldsymbol{\varphi}^0, \boldsymbol{\sigma}^2) \times p(\boldsymbol{w} \mid \boldsymbol{\varphi}, X).
\]
Again, looking to isolate the terms relating to $\boldsymbol{\varphi}_{k}$, we get: 
\begin{align}
    p(\boldsymbol{\varphi} \mid \boldsymbol{\varphi}^0, \boldsymbol{\sigma}^2) &\propto f(\boldsymbol{\varphi}_{k}; \boldsymbol{\varphi}^0_k, \sigma_k^2I_P) \\
    p(\boldsymbol{w} \mid \boldsymbol{\varphi}, X) &= \prod_{i=1}^N \prod_{s=1}^\infty \left[\Phi(\boldsymbol{x}_i^{\intercal}\boldsymbol{\varphi}_s)\prod_{r=1}^{s-1}\left(1 - \Phi(\boldsymbol{x}_i^{\intercal}\boldsymbol{\varphi}_r\right)\right]^{\mathbb{I}\{w_i = s\}}\\ 
    &\propto \prod_{i=1}^N\prod_{s=k}^\infty \left[\Phi(\boldsymbol{x}_i^{\intercal}\boldsymbol{\varphi}_s)\prod_{r=1}^{s-1}\left(1 - \Phi(\boldsymbol{x}_i^{\intercal}\boldsymbol{\varphi}_r)\right)\right]^{\mathbb{I}\{w_i = s\}} \\
    &\propto \prod_{i=1}^N\left[\Phi(\boldsymbol{x}_i^{\intercal}\boldsymbol{\varphi}_k)^{\mathbb{I}\{w_i = k\}}\prod_{s=k+1}^\infty (1 - \Phi(\boldsymbol{x}_i^{\intercal}\boldsymbol{\varphi}_k))^{\mathbb{I}\{w_i = s\}}\right].
\end{align}
Putting these terms together, the complete conditional distribution for $\boldsymbol{\varphi}_k$ takes the form:
\begin{align}
    p(\boldsymbol{\varphi}_{k} \mid \Theta_{-\boldsymbol{\varphi_{k}}}, A, X) &\propto f(\boldsymbol{\varphi}_{k}; \boldsymbol{\varphi}^0_k, \sigma_k^2I_P) \times \prod_{i=1}^N\left[\Phi(\boldsymbol{x}_i^{\intercal}\boldsymbol{\varphi}_k)^{\mathbb{I}\{w_i = k\}}\prod_{s=k+1}^\infty (1 - \Phi(\boldsymbol{x}_i^{\intercal}\boldsymbol{\varphi}_k))^{\mathbb{I}\{w_i = s\}}\right].
    \label{eqn:complete_conditional_phi}
\end{align}

\subsection{Complete conditional distribution for \texorpdfstring{$\boldsymbol{\varphi}^0_k$}{phi0}}
\label{app:complete_conditional_phi0}

The complete conditional distribution for $\boldsymbol{\varphi}^0_k$ is written as 
\[
p(\boldsymbol{\varphi}^0_k \mid \Theta_{-\boldsymbol{\varphi}^0_k}, A, X) \propto p(\boldsymbol{\varphi} \mid \boldsymbol{\varphi}^0, \boldsymbol{\sigma}^2) \times p(\boldsymbol{\varphi}^0).
\]
Isolating terms involving $\boldsymbol{\varphi}^0_k$ gives
\begin{align}
    p(\boldsymbol{\varphi}^0_k \mid \Theta_{-\boldsymbol{\varphi}^0_k}, A, X) \propto p(\boldsymbol{\varphi}_k \mid \boldsymbol{\varphi}^0_k, \boldsymbol{\sigma}_k^2) \times p(\boldsymbol{\varphi}_k^0) \propto  f(\boldsymbol{\varphi}_k;\boldsymbol{\varphi}^0_k, \sigma_k^2I_P) \times f(\boldsymbol{\varphi}_k^0; \boldsymbol{\mu}, I_P).
\end{align}
Replacing $f(\cdot)$ with the corresponding density functions gives: 
\begin{align}
    f(\boldsymbol{\varphi}_k;\boldsymbol{\varphi}^0_k, \sigma_k^2I_P)\times f(\boldsymbol{\varphi}_k^0; \boldsymbol{\mu}, I_P) \propto \exp\left\{-\frac{1}{2}\left(\boldsymbol{\varphi}_k^0 - \frac{\boldsymbol{\varphi}_k + \sigma_k^2\boldsymbol{\mu}}{\sigma_k^2 + 1}\right)^{\intercal}\left(\frac{1}{\sigma_k^2} + 1\right)I_P\left(\boldsymbol{\varphi}_k^0 - \frac{\boldsymbol{\varphi}_k + \sigma_k^2\boldsymbol{\mu}}{\sigma_k^2 + 1}\right)\right\}.
    \label{eqn:complete_conditional_phi_0}
\end{align}
It then follows that:
\begin{equation}
    \boldsymbol{\varphi}_k^0 \mid \Theta_{-\boldsymbol{\varphi}_k^0}, A, X \sim \text{Normal}\left(\frac{\boldsymbol{\varphi}_k + \sigma_k^2\boldsymbol{\mu}}{\sigma_k^2 + 1}, \frac{\sigma_k^2}{\sigma_k^2 + 1}I_P\right).
\end{equation}

\subsection{Complete conditional distribution for \texorpdfstring{$\gamma_{ks}^{\prime}$}{gamma}}
\label{app:complete_conditional_gammaks}

The complete conditional distribution can be  written as:
\begin{equation}
p(\gamma_{ks}^{\prime} \mid \Theta_{-\boldsymbol{\gamma_{ks}^{\prime}}}, A, X) \propto p(z \mid \boldsymbol{w}, \boldsymbol{\gamma}^{\prime}) \times p(\boldsymbol{\gamma}^{\prime})
\end{equation}
Isolating the terms with $\gamma_{ks}^{\prime}$, we obtain:
\begin{align}
    p(z\mid \boldsymbol{w}, \boldsymbol{\gamma}^{\prime}) &= \prod_{\ell=1}^L \prod_{i=1}^N \prod_{w,r=1}^\infty \left[\gamma_{wr}^{\prime}\prod_{m=1}^{r-1}\left(1 - \gamma_{wm}^{\prime}\right)\right]^{\mathbb{I}\{w_i = w,z_{\ell i } = m\}} \\
    &\propto \prod_{\ell=1}^L \prod_{i=1}^N \left[\left(\gamma_{ks}^{\prime}\right)^{\mathbb{I}\{z_{\ell i} = s\}}\prod_{r=s+1}^{\infty}\left(1 - \gamma_{ks}^{\prime}\right)^{\mathbb{I}\{z_{\ell i} = r\}}\right]^{\mathbb{I}\{w_i = k\}}, \\
    p(\boldsymbol{\gamma}^{\prime}) &\propto g(\gamma_{ks}^{\prime};1,\eta_0).
\end{align}
Combining these terms, we obtain the complete conditional in the form:
\begin{align}
    \gamma_{ks}^{\prime} \mid \Theta_{-\boldsymbol{\gamma}_{ks}^{\prime}}, A, X \sim \text{Beta}\left(\sum_{\ell = 1}^L \sum_{i=1}^N \mathbb{I}\{z_{\ell i} = s, w_i = k\} + 1, \eta_0 + \sum_{r=s+1}^\infty \sum_{\ell = 1}^L \sum_{i=1}^N \mathbb{I}\{z_{\ell i} = r, w_i = k\}\right).
    \label{eqn:complete_conditional_gamma}
\end{align}

\subsection{Complete conditional distribution for \texorpdfstring{$\rho_{km}$}{rho}}
\label{app:complete_conditional_rhokm}

The complete conditional distribution takes the form:
\begin{align}
    p(\rho_{km} \mid \Theta_{-\rho_{km}}, A, X) \propto p(A \mid z, \rho) \times p(\rho)
\end{align}
We have
\begin{align}
    p(A \mid z, \rho) &\propto \prod_{\ell=1}^L \prod_{i,j\in [N]\setminus\{i\}} \left[\rho_{km}^{A_{\ell i j}}(1 - \rho_{km})^{1 - A_{\ell i j}}\right]^{\mathbb{I}\{z_{\ell i} = k, z_{\ell j} = m\}},
\end{align}
which gives: 
\begin{align}
    p(\rho_{km}\mid\Theta_{-\rho_{km}}, A, X) \propto \prod_{\ell=1}^L \prod_{i,j\in [N], i\neq j}\left[\rho_{km}^{A_{\ell i j}}(1 - \rho_{km})^{1 - A_{\ell i j}}\right]^{\mathbb{I}\{z_{\ell i} = k, z_{\ell j} = m\}} \times \rho_{km}^{\alpha_0 - 1} (1 - \rho_{km})^{\beta_0 - 1}.
\end{align}
It follows that: 
\begin{align}
    \rho_{km} \mid \Theta_{-\rho_{km}}, A, X \sim \text{Beta}\Bigg(&\alpha_0 + \sum_{\ell=1}^L\sum_{i,j\in[N], i\neq j} A_{\ell ij}\mathbb{I}\{z_{\ell i} = k, z_{\ell j} = m\}, \\
    &\beta_0 + \sum_{\ell=1}^L\sum_{i,j\in[N], i\neq j} (1 - A_{\ell ij})\mathbb{I}\{z_{\ell i} = k, z_{\ell j} = m\}\Bigg).
    \label{eqn:complete_conditional_rho}
\end{align}




\subsection{Complete conditional distribution for \texorpdfstring{$\sigma_k^2$}{sigma}}
\label{app:complete_conditional_sigmak2}

The complete conditional distribution for $\sigma_k^2$ takes the form:
\begin{equation}
p(\sigma_k^2 \mid \Theta_{-\sigma^2_k}, A, X) \propto p(\boldsymbol{\varphi} \mid \boldsymbol{\varphi}^0, \boldsymbol{\sigma}^2) \times p(\boldsymbol{\sigma}^2).
\end{equation}
Isolating relevant terms, we obtain:
\begin{align}
    p(\sigma_k^2 \mid \Theta_{-\sigma^2_k}, A, X) &\propto p(\boldsymbol{\varphi}_k \mid \boldsymbol{\varphi}_k^0, \sigma_k^2) \times p(\sigma_{k}^2).
\end{align}
Therefore:
\begin{align}
    p(\sigma_k^2 \mid \Theta_{-\sigma_k^2}, A, X) &\propto \left(\sigma_k^2\right)^{-(P/2 + \nu_0 + 1)}\exp\left\{-\frac{1}{\sigma_k^2}\left[\omega_0 + \frac{1}{2}\left(\boldsymbol{\varphi}_k - \boldsymbol{\varphi}_k^0\right)^{\intercal}\left(\boldsymbol{\varphi}_k - \boldsymbol{
        \varphi}_k^0\right)\right]\right\}.
\end{align}
It then follows that:
\begin{align}
    \sigma_k^2 \mid \Theta_{-\sigma_k^2}, A, X \sim \text{Inverse\text{-}Gamma}\left(\nu_0 + \frac{P}{2}, \omega_0 + \frac{1}{2}\left(\boldsymbol{\varphi}_k - \boldsymbol{\varphi}_k^0\right)^{\intercal}\left(\boldsymbol{\varphi}_k - \boldsymbol{
        \varphi}_k^0\right)\right).
\end{align}

\section{Gradients for the ELBO with respect to \texorpdfstring{$q(\boldsymbol{\varphi})$}{phi0}}
\label{app:gradients}

We consider all variational forms for parameters other than $\boldsymbol{\varphi}_m$ as fixed. This avoids back propagation and keeps the update tractable. If we isolate terms of the ELBO that contain $\Tilde{\boldsymbol{\theta}}_m$ and $\widetilde{\Sigma}_m$, we obtain 
\begin{align}
    \mathcal{L} &:= \mathbb{E}_q\left\{\log p(\Theta, A \mid X)\right\} - \mathbb{E}\left\{\log q(\Theta)\right\} \\
    &= \mathbb{E}_q\left\{\log p(\boldsymbol{w} \mid \boldsymbol{\varphi}, X) + \log p(\boldsymbol{\varphi}\mid \boldsymbol{\varphi}^0, \boldsymbol{\sigma}^2) - \log q(\boldsymbol{\varphi})\right\} + \kappa \\
    &= \mathbb{E}_q\left\{\sum_{i=1}^N (\boldsymbol{w}_i)_m\log \left[\Phi\left(\boldsymbol{x}_i^{\intercal}\boldsymbol{\varphi}_m\right)\right] + \sum_{i=1}^N \sum_{k=m+1}^\infty (\boldsymbol{w}_i)_k\log \left[1 - \Phi\left(\boldsymbol{x}_i^{\intercal}\boldsymbol{\varphi}_m\right)\right]\right.\\
    &\left.-\frac{1}{2\sigma_m^2}\left(\boldsymbol{\varphi}_m - \boldsymbol{\varphi}_m^0\right)^{\intercal}\left(\boldsymbol{\varphi}_m - \boldsymbol{\varphi}_m^0\right) + \frac{1}{2}\log\left[\text{det}\left(\widetilde{\Sigma}_m\right)\right] + \frac{1}{2}\left(\boldsymbol{\varphi}_m - \Tilde{\boldsymbol{\theta}}_m\right)^{\intercal}\widetilde{\Sigma}_m^{-1}\left(\boldsymbol{\varphi}_m - \Tilde{\boldsymbol{\theta}}_m\right)\right\} + \kappa \\
    &= \sum_{i=1}^N (\Tilde{\boldsymbol{\phi}}_{w_i})_m\mathbb{E}_q\left\{\log \left[\Phi\left(\boldsymbol{x}_i^{\intercal}\boldsymbol{\varphi}_m\right)\right]\right\} + \sum_{i=1}^N \sum_{k=m+1}^\infty (\Tilde{\boldsymbol{\phi}}_{w_i})_k\mathbb{E}_q\left\{\log \left[1 - \Phi\left(\boldsymbol{x}_i^{\intercal}\boldsymbol{\varphi}_m\right)\right]\right\}\\
    &-\frac{\nu_m}{2\omega_m}\left[\text{tr}\left(\widetilde{\Sigma}_m\right) + \Tilde{\boldsymbol{\theta}}_m^{\intercal}\Tilde{\boldsymbol{\theta}}_m - 2\Tilde{\boldsymbol{\theta}}_m^{\intercal}\Tilde{\boldsymbol{\theta}}_m^0\right] + \frac{1}{2}\log\left[\text{det}\left(\widetilde{\Sigma}_m\right)\right] + \kappa,
\end{align}
where in the final stage, we use that if $A$ is some square matrix and $\boldsymbol{Y} \sim \text{Normal}(\boldsymbol{\mu}, \Sigma)$, then $\mathbb{E}\left\{\boldsymbol{Y}^{\intercal}A\boldsymbol{Y}\right\} = \text{tr}(A\Sigma) + \boldsymbol{\mu}^{\intercal}A\boldsymbol{\mu}$, which yields that $\mathbb{E}_{q}\left\{(\boldsymbol{\varphi}_m - \Tilde{\boldsymbol{\theta}}_m)^{\intercal}\widetilde{\Sigma}_m^{-1}(\boldsymbol{\varphi}_m - \Tilde{\boldsymbol{\theta}}_m)\right\} = \text{tr}\left(\widetilde{\Sigma}_m^{-1}\widetilde{\Sigma}_m\right) = P$, where $P$ is the length of the vector $\Tilde{\boldsymbol{\theta}}_m$. We use $\kappa$ to denote a constant. 

To compute the gradients, we first note the following
\begin{align}
    \pdv{\Tilde{\boldsymbol{\theta}}_m}\left\{f(\boldsymbol{\varphi}_m; \Tilde{\boldsymbol{\theta}}_m, \widetilde{\Sigma}_m)\right\} &=\widetilde{\Sigma}_m^{-1}(\boldsymbol{\varphi}_m - \Tilde{\boldsymbol{\theta}}_m) \times f(\boldsymbol{\varphi}_m; \Tilde{\boldsymbol{\theta}}_m, \widetilde{\Sigma}_m), \\
    \pdv{\widetilde{\Sigma}_m}\left\{f(\boldsymbol{\varphi}_m; \Tilde{\boldsymbol{\theta}}_m, \widetilde{\Sigma}_m)\right\} &= \frac{1}{2}\left(\widetilde{\Sigma}_m^{-1}(\boldsymbol{\varphi}_m - \Tilde{\boldsymbol{\theta}}_m)(\boldsymbol{\varphi}_m - \Tilde{\boldsymbol{\theta}}_m)^{\intercal}\widetilde{\Sigma}_m^{-1} -\widetilde{\Sigma}_m^{-1}\right) \times f(\boldsymbol{\varphi}_m; \Tilde{\boldsymbol{\theta}}_m, \widetilde{\Sigma}_m).
\end{align}
Exchaning integration and differentiation, we then find
\begin{align}
    \pdv{\Tilde{\boldsymbol{\theta}}_m} \mathbb{E}_q\left\{\log \left[\Phi\left(\boldsymbol{x}_i^{\intercal}\boldsymbol{\varphi}_m\right)\right]\right\} &= \pdv{\Tilde{\boldsymbol{\theta}}_m}\int_{\mathbb{R}^P} \log\left[\Phi\left(\boldsymbol{x}_i^{\intercal}\boldsymbol{\varphi}_m\right)\right] f(\boldsymbol{\varphi}_m; \Tilde{\boldsymbol{\theta}}_m, \widetilde{\Sigma}_m) \dd \boldsymbol{\varphi}_m \\
    &= \int_{\mathbb{R}^P}\widetilde{\Sigma}_m^{-1}(\boldsymbol{\varphi}_m - \Tilde{\boldsymbol{\theta}}_m)\log\left[\Phi\left(\boldsymbol{x}_i^{\intercal}\boldsymbol{\varphi}_m\right)\right]  f(\boldsymbol{\varphi}_m; \Tilde{\boldsymbol{\theta}}_m, \widetilde{\Sigma}_m) \\
    &= \mathbb{E}_q\left\{\widetilde{\Sigma}_m^{-1}(\boldsymbol{\varphi}_m - \Tilde{\boldsymbol{\theta}}_m) \log\left[\Phi\left(\boldsymbol{x}_i^{\intercal}\boldsymbol{\varphi}_m\right)\right]\right\}, \\
    \pdv{\widetilde{\Sigma}_m}\mathbb{E}_q\left\{\log \left[\Phi\left(\boldsymbol{x}_i^{\intercal}\boldsymbol{\varphi}_m\right)\right]\right\} &= \pdv{\widetilde{\Sigma}_m}\int_{\mathbb{R}^P} \log\left[\Phi\left(\boldsymbol{x}_i^{\intercal}\boldsymbol{\varphi}_m\right)\right] f(\boldsymbol{\varphi}_m; \Tilde{\boldsymbol{\theta}}_m, \widetilde{\Sigma}_m) \dd \boldsymbol{\varphi}_m \\
    &= \int_{\mathbb{R}^P} \frac{1}{2}\left(\widetilde{\Sigma}_m^{-1}(\boldsymbol{\varphi}_m - \Tilde{\boldsymbol{\theta}}_m)(\boldsymbol{\varphi}_m - \Tilde{\boldsymbol{\theta}}_m)^{\intercal}\widetilde{\Sigma}_m^{-1} -\widetilde{\Sigma}_m^{-1}\right)\\
    &\qquad \times\log\left[\Phi\left(\boldsymbol{x}_i^{\intercal}\boldsymbol{\varphi}_m\right)\right] f(\boldsymbol{\varphi}_m; \Tilde{\boldsymbol{\theta}}_m, \widetilde{\Sigma}_m) \dd \boldsymbol{\varphi}_m \\
    &= \mathbb{E}_q\left\{\frac{1}{2}\left(\widetilde{\Sigma}_m^{-1}(\boldsymbol{\varphi}_m - \Tilde{\boldsymbol{\theta}}_m)(\boldsymbol{\varphi}_m - \Tilde{\boldsymbol{\theta}}_m)^{\intercal}\widetilde{\Sigma}_m^{-1} -\widetilde{\Sigma}_m^{-1}\right)\log\left[\Phi\left(\boldsymbol{x}_i^{\intercal}\boldsymbol{\varphi}_m\right)\right]\right\},
\end{align}
with analogous results holding for derivatives of $\mathbb{E}_q\left\{\log \left[1 - \Phi\left(\boldsymbol{x}_i^{\intercal}\boldsymbol{\varphi}_m\right)\right]\right\}$. Using these results, we obtain the derivatives of the ELBO as:
\begin{align}
    \pdv{\mathcal{L}}{\Tilde{\boldsymbol{\theta}}_m} =& \,\sum_{i=1}^N (\Tilde{\boldsymbol{\phi}}_{w_i})_m\mathbb{E}_q\left\{\widetilde{\Sigma}_m^{-1}(\boldsymbol{\varphi}_m - \Tilde{\boldsymbol{\theta}}_m) \log\left[\Phi\left(\boldsymbol{x}_i^{\intercal}\boldsymbol{\varphi}_m\right)\right]\right\}\\ 
    &+ \sum_{i=1}^N \sum_{k=m+1}^\infty (\Tilde{\boldsymbol{\phi}}_{w_i})_k\mathbb{E}_q\left\{\widetilde{\Sigma}_m^{-1}(\boldsymbol{\varphi}_m - \Tilde{\boldsymbol{\theta}}_m) \log\left[1 - \Phi\left(\boldsymbol{x}_i^{\intercal}\boldsymbol{\varphi}_m\right)\right]\right\} - \frac{\Tilde{\nu}_m}{\Tilde{\omega}_m}\left(\Tilde{\boldsymbol{\theta}}_m - \Tilde{\boldsymbol{\theta}}_m^0\right)\\
    \pdv{\mathcal{L}}{\widetilde{\Sigma}_m} =& \,\sum_{i=1}^N (\Tilde{\boldsymbol{\phi}}_{w_i})_m\mathbb{E}_q\left\{\frac{1}{2}\left(\widetilde{\Sigma}_m^{-1}(\boldsymbol{\varphi}_m - \Tilde{\boldsymbol{\theta}}_m)(\boldsymbol{\varphi}_m - \Tilde{\boldsymbol{\theta}}_m)^{\intercal}\widetilde{\Sigma}_m^{-1} -\widetilde{\Sigma}_m^{-1}\right)\log\left[\Phi\left(\boldsymbol{x}_i^{\intercal}\boldsymbol{\varphi}_m\right)\right]\right\}\\ 
    &+ \sum_{i=1}^N \sum_{k=m+1}^\infty (\Tilde{\boldsymbol{\phi}}_{w_i})_k\mathbb{E}_q\left\{\frac{1}{2}\left(\widetilde{\Sigma}_m^{-1}(\boldsymbol{\varphi}_m - \Tilde{\boldsymbol{\theta}}_m)(\boldsymbol{\varphi}_m - \Tilde{\boldsymbol{\theta}}_m)^{\intercal}\widetilde{\Sigma}_m^{-1} -\widetilde{\Sigma}_m^{-1}\right)\log\left[1 - \Phi\left(\boldsymbol{x}_i^{\intercal}\boldsymbol{\varphi}_m\right)\right]\right\}\\
    &- \frac{\nu_k}{2\omega_k}I_P + \frac{1}{2}\widetilde{\Sigma}_m^{-1}.
\end{align}

\section{ELBO}
\label{app:ELBO}
The logarithm of the joint distribution for the hierarchical multiplex stochastic block model takes the form:
\begin{align}
\log p(\Theta, A, X) =&  \,
\sum_{\ell \in \mathcal{L}} \sum_{i,j\in\mathcal{V}, i\neq j} \sum_{k,m=1}^\infty \mathbb{I}\{z_{\ell i} = k, z_{\ell j} = m\}\big[A_{\ell ij}\log \rho_{km} + (1 - A_{\ell ij})\log(1 - \rho_{km})\big] \\
&+ \sum_{k,m=1}^\infty \big[(\alpha_0 - 1)\log \rho_{km} + (\beta_0 - 1)\log(1 - \rho_{km})\big] \\
&+ \sum_{\ell \in \mathcal{L}}\sum_{i \in \mathcal{V}} \sum_{k=1}^\infty \sum_{w=1}^\infty \mathbb{I}\{z_{\ell i} = k, w_i = w\}\left[\log \gamma_{wk}^{\prime} + \sum_{r=1}^{k-1}\log(1 - \gamma_{wr}^{\prime})\right] \\
&+ \sum_{i\in\mathcal{V}} \sum_{w=1}^\infty \mathbb{I}\{w_i = w\}\left[\log \Phi(\boldsymbol{x}_i^{\intercal}\boldsymbol{\varphi}_w) + \sum_{k=1}^{w-1}\log(1 - \Phi(\boldsymbol{x}_i^{\intercal}\boldsymbol{\varphi}_k))\right] \\
&+ \sum_{k=1}^\infty \left[-\frac{P}{2}\log\sigma_k^2 - \frac{1}{2\sigma_k^2}(\boldsymbol{\varphi}_k - \boldsymbol{\varphi}_k^0)^{\intercal}(\boldsymbol{\varphi}_k - \boldsymbol{\varphi}_k^0)\right] \\
&+ \sum_{k=1}^\infty \left[-\frac{1}{2}(\boldsymbol{\varphi}_k^0 - \boldsymbol{\mu})^{\intercal}(\boldsymbol{\varphi}_k^0 - \boldsymbol{\mu})\right]\\
&+ \sum_{k=1}^\infty \left[-(\nu_0 + 1)\log\sigma_k^2 - \frac{\omega_0}{\sigma_k^2}\right] \\
&+ \sum_{k=1}^\infty \sum_{s=1}^\infty \left[(\eta_0 - 1)\log(1 - \gamma_{ks}^{\prime})\right] + \kappa.
 \end{align}
After computing the expectations, we obtain:
\begin{align}
\mathbb{E}_q&\{\log p(\boldsymbol{\theta}, A, X)\} \\    
=& \sum_{\ell \in \mathcal{L}} \sum_{i,j\in\mathcal{V}, i\neq j} \sum_{k,m=1}^\infty (\Tilde{\boldsymbol{\phi}}_{z_{\ell i}})_k(\Tilde{\boldsymbol{\phi}}_{z_{\ell j}})_m\bigg[A_{\ell ij}\left\{\psi(\Tilde{\alpha}_{\rho_{km}}) - \psi(\Tilde{\alpha}_{\rho_{km}} + \Tilde{\beta}_{\rho_{km}})\right\} \\&+ (1 - A_{\ell ij})\left\{\psi(\Tilde{\beta}_{\rho_{km}}) - \psi(\Tilde{\alpha}_{\rho_{km}} + \Tilde{\beta}_{\rho_{km}})\right\}\bigg] \\
&+ \sum_{k,m=1}^\infty \bigg[(\alpha_0 - 1)\left\{\psi(\Tilde{\alpha}_{\rho_{km}}) - \psi(\Tilde{\alpha}_{\rho_{km}} + \Tilde{\beta}_{\rho_{km}})\right\} + (\beta_0 - 1)\left\{\psi(\Tilde{\beta}_{\rho_{km}}) - \psi(\Tilde{\alpha}_{\rho_{km}} + \Tilde{\beta}_{\rho_{km}})\right\}\bigg] \\
&+ \sum_{\ell\in\mathcal{L}}\sum_{i\in\mathcal{V}}\sum_{k=1}^\infty \sum_{w=1}^\infty (\Tilde{\boldsymbol{\phi}}_{z_{\ell i}})_k(\Tilde{\boldsymbol{\phi}}_{w_{i}})_w\bigg[\psi(\Tilde{\alpha}_{\gamma_{wk}}) - \psi(\Tilde{\alpha}_{\gamma_{wk}} + \Tilde{\beta}_{\gamma_{wk}}) + \sum_{r=1}^{k-1}\left\{\psi(\Tilde{\beta}_{\gamma_{wr}}) - \psi(\Tilde{\alpha}_{\gamma_{wr}} + \Tilde{\beta}_{\gamma_{wr}})\right\}\bigg] \\
&+ \sum_{i\in\mathcal{V}} \sum_{w=1}^\infty (\Tilde{\boldsymbol{\phi}}_{w_i})_w\big[\mathbb{E}_q\left\{\log \Phi(\boldsymbol{x}_i^{\intercal}\boldsymbol{\varphi}_w)\right\} + \sum_{k=1}^{w-1}\mathbb{E}_q\left\{\log(1 - \Phi(\boldsymbol{x}_i^{\intercal}\boldsymbol{\varphi}_k))\right\}\big] \\
&+ \sum_{k=1}^\infty \left[-\frac{P}{2}\left\{\log \Tilde{\omega}_k - \psi(\Tilde{\nu}_k)\right\} - \frac{\Tilde{\nu}_k}{2\Tilde{\omega}_k}\left\{(\boldsymbol{\theta}_{\varphi_k} - \boldsymbol{\theta}_{\varphi_k^0})^{\intercal}(\boldsymbol{\theta}_{\varphi_k} - \boldsymbol{\theta}_{\varphi_k^0}) + \text{tr}\left(\widetilde{\Sigma}_{\varphi_k}\right)+ \text{tr}\left(\widetilde{\Sigma}_{\varphi_k^0}\right)\right\}\right] \\
&+ \sum_{k=1}^\infty \left[-\frac{1}{2}\left\{(\boldsymbol{\theta}_{\varphi_k^0} - \boldsymbol{\mu})^{\intercal}(\boldsymbol{\theta}_{\varphi_k^0} - \boldsymbol{\mu}) + \text{tr}\left(\widetilde{\Sigma}_{\varphi_k^0}\right)\right\}\right]\\
&+ \sum_{k=1}^\infty \left[-(\nu_0 + 1)\left\{\log \Tilde{\omega}_k - \psi(\Tilde{\nu}_k)\right\} - \frac{\omega_0\Tilde{\nu}_k}{\Tilde{\omega}_k}\right] \\
&+ \sum_{k=1}^\infty \sum_{s=1}^\infty \left[(\eta_0 - 1)\left\{\psi(\Tilde{\beta}_{\gamma_{ks}}) - \psi(\Tilde{\alpha}_{\gamma_{ks}} + \Tilde{\beta}_{\gamma_{ks}})\right\}\right]  + \kappa.
\end{align}
The logarithm of the variational approximation $q(\Theta)$ takes the form:
\begin{align}
\log q(\Theta) =& \, \sum_{\ell \in \mathcal{L}}\sum_{i \in \mathcal{V}} \sum_{k=1}^{M_u}\mathbb{I}\{z_{\ell i} = k\}\log\left\{(\Tilde{\boldsymbol{\phi}}_{z_{\ell i}})_k\right\} \\
&+ \sum_{k,m=1}^{M_w} \bigg[\log\Gamma(\Tilde{\alpha}_{\rho_{km}} + \Tilde{\beta}_{\rho_{km}}) - \log \Gamma(\Tilde{\alpha}_{\rho_{km}}) - \log \Gamma(\Tilde{\beta}_{\rho_{km}})\\ 
&+ (\Tilde{\alpha}_{km} - 1) \log \rho_{km} + (\Tilde{\beta}_{km} - 1)\log(1 - \rho_{km})\bigg] + \sum_{i\in\mathcal{V}}\sum_{w=1}^{M_w} \mathbb{I}\{w_i = w\}\log\left\{(\Tilde{\boldsymbol{\phi}}_{w_i})_w\right\} \\
&+ \sum_{k=1}^{M_w} \sum_{s=1}^{M_u}\bigg[\log\Gamma(\Tilde{\alpha}_{\gamma_{ks}} + \Tilde{\beta}_{\gamma_{ks}}) - \log \Gamma(\Tilde{\alpha}_{\gamma_{ks}}) - \log \Gamma(\Tilde{\beta}_{\gamma_{ks}}) \\
&+ (\Tilde{\alpha
}_{\gamma_{ks}} - 1)\log\gamma_{ks}^{\prime} + (\Tilde{\beta}_{\gamma_{ks}} - 1)\log(1 - \gamma_{ks}^{\prime})\bigg] \\
&+ \sum_{k=1}^{M_w} \left[-\frac{1}{2}\log \text{det}\left(\Tilde{\Sigma}_{\varphi_k}\right) - \frac{1}{2}\left(\boldsymbol{\varphi}_k - \Tilde{\boldsymbol{\theta}}_{\varphi_k}\right)^{\intercal}\Tilde{\Sigma}_{\varphi_k}^{-1}\left(\boldsymbol{\varphi}_k - \Tilde{\boldsymbol{\theta}}_{\varphi_k}\right)\right] \\
&+ \sum_{k=1}^{M_w}  \left[-\frac{1}{2}\log \text{det}\left(\Tilde{\Sigma}_{\varphi_k^0}\right) - \frac{1}{2}\left(\boldsymbol{\varphi}_k^0 - \Tilde{\boldsymbol{\theta}}_{\varphi_k^0}\right)^{\intercal}\Tilde{\Sigma}_{\varphi_k^0}^{-1}\left(\boldsymbol{\varphi}_k - \Tilde{\boldsymbol{\theta}}_{\varphi_k^0}\right)\right] \\
&+ \sum_{k=1}^{M_w}\left[\Tilde{\nu}_k\log \Tilde{\omega}_k - \log \Gamma(\Tilde{\nu}_k) -(\Tilde{\nu}_k + 1) \log \sigma_k^2 - \frac{
\Tilde{\omega}_k}{\sigma_k^2}\right]  + \kappa.
\end{align}
We can compute the expectations to yield:
\begin{align}
\mathbb{E}_q&\left\{\log q(\Theta)\right\} = \sum_{\ell \in \mathcal{L}}\sum_{i \in \mathcal{V}} \sum_{k=1}^{M_u}(\Tilde{\boldsymbol{\phi}}_{z_{\ell i}})_k\log\left\{(\Tilde{\boldsymbol{\phi}}_{z_{\ell i}})_k\right\} \\
&+ \sum_{k,m=1}^{M_w} \bigg[\log\Gamma(\Tilde{\alpha}_{\rho_{km}} + \Tilde{\beta}_{\rho_{km}}) - \log \Gamma(\Tilde{\alpha}_{\rho_{km}}) - \log \Gamma(\Tilde{\beta}_{\rho_{km}})\\ 
&+ (\Tilde{\alpha}_{km} - 1) \left\{\psi(\Tilde{\alpha}_{\rho_{km}}) - \psi(\Tilde{\alpha}_{\rho_{km}} + \Tilde{\beta}_{\rho_{km}})\right\} + (\Tilde{\beta}_{km} - 1)\left\{\psi(\Tilde{\beta}_{\rho_{km}}) - \psi(\Tilde{\alpha}_{\rho_{km}} + \Tilde{\beta}_{\rho_{km}})\right\}\bigg] \\
&+ \sum_{i\in\mathcal{V}}\sum_{w=1}^{M_w} (\Tilde{\boldsymbol{\phi}}_{w_{i}})_w\log\left\{(\Tilde{\boldsymbol{\phi}}_{w_i})_w\right\} + \sum_{k=1}^{M_w} \sum_{s=1}^{M_u}\bigg[\log\Gamma(\Tilde{\alpha}_{\gamma_{ks}} + \Tilde{\beta}_{\gamma_{ks}}) - \log \Gamma(\Tilde{\alpha}_{\gamma_{ks}}) - \log \Gamma(\Tilde{\beta}_{\gamma_{ks}}) \\
&+ (\Tilde{\alpha
}_{\gamma_{ks}} - 1)\left\{\psi(\Tilde{\alpha}_{\gamma_{ks}}) - \psi(\Tilde{\alpha}_{\gamma_{ks}} + \Tilde{\beta}_{\gamma_{ks}})\right\} + (\Tilde{\beta}_{\gamma_{ks}} - 1)\left\{\psi(\Tilde{\beta}_{\gamma_{ks}}) - \psi(\Tilde{\alpha}_{\gamma_{ks}} + \Tilde{\beta}_{\gamma_{ks}})\right\}\bigg] \\
&+ \sum_{k=1}^{M_w} \left[-\frac{1}{2}\log \text{det}\left(\Tilde{\Sigma}_{\varphi_k}\right)\right] + \sum_{k=1}^{M_w}  \left[-\frac{1}{2}\log \text{det}\left(\Tilde{\Sigma}_{\varphi_k^0}\right)\right] \\
&+ \sum_{k=1}^{M_w}\left[(\Tilde{\nu}_k + 1)\psi(\Tilde{\nu}_k) -\log \Tilde{\omega}_k - \log \Gamma(\Tilde{\nu}_k) - \Tilde{\nu}_k\right]  + \kappa.
\end{align}
These two expressions can then be combined to obtain the ELBO.

\section{Full algorithm of CAVI with Adam}
\label{app:full_algorithm}

\begin{breakablealgorithm}
\caption{\textit{CAVI with Adam}. Write $f_k(\boldsymbol{\theta}, \Sigma) = \partial \mathcal{L} / \partial \Tilde{\boldsymbol{\theta}}_{\varphi_k}$ evaluated at $\Tilde{\boldsymbol{\theta}}_{\varphi_k} = \boldsymbol{\theta}$ and $\widetilde{{\Sigma}}_{\varphi_k} = \Sigma$ conditional upon all other current parameter estimates, and similarly $g_k(\boldsymbol{\theta}, \Sigma) = \partial \mathcal{L} / \partial \widetilde{\Sigma}_{\varphi_k}$, as in \eqref{eqn:variational_theta}-\eqref{eqn:variational_sigma}. Write $\mathcal{L}(\boldsymbol{\theta},\Sigma)$ for the ELBO evaluated at $\Tilde{\boldsymbol{\theta}}_{\varphi_k} = \boldsymbol{\theta}$ and $\widetilde{\Sigma}_{\varphi_k} = \Sigma$.}
\begin{algorithmic}[1]
\item Select values for $\eta_1, \eta_1$, and $\beta_1, \beta_2 \in [0,1)$ and $\epsilon$. Provide maximum number of decreases $T$ and maximum number of steps $S$. 
\item Initialise $b_{1}, c_{1}\gets \{0\}^{M_w \times P}$, $b_{2}, c_{2}\gets \{0\}^{M_w\times P\times P}$ and $s_1, s_2 \gets \{0\}^{M_w}$. \item Initialise $\boldsymbol{\theta}^{\text{best}}, \boldsymbol{\theta}^{\text{curr}} = \{0\}^{M_w \times P}$ and $\Sigma^{\text{best}}, \Sigma^{\text{curr}} = \{0\}^{M_w \times P \times P}$. 
\While {ELBO for $\Theta$ not converged}
\item Update $\hat{q}(\rho_{km})$ for $k,m \in [M_w]$ using \eqref{eqn:variational_alpha_rho}-\eqref{eqn:variational_beta_rho}.
\item Update $\hat{q}(\gamma_{km})$ for $k\in [M_w], m \in [M_z]$ using \eqref{eqn:variational_alpha_gamma}-\eqref{eqn:variational_beta_gamma}.
\item Update $\hat{q}(\boldsymbol{\varphi}_k^0)$ for $k\in [M_w]$ using \eqref{eqn:variational_theta0}-\eqref{eqn:variational_sigma0}.
\For {$k=1$ to $M_w$} 
    \State $\boldsymbol{\theta}^{\text{best}}[k] \gets \Tilde{\boldsymbol{\theta}}_{\varphi_k}$ and $\boldsymbol{\theta}^{\text{curr}}[k] \gets \Tilde{\boldsymbol{\theta}}_{\varphi_k}$.
    \State $\Sigma^{\text{best}}[k] \gets \widetilde{\Sigma}_{\varphi_k}$ and $\Sigma^{\text{curr}}[k] \gets \widetilde{\Sigma}_{\varphi_k}$.
\EndFor
\State $\mathcal{L}^{\text{best}} \gets \mathcal{L}(\boldsymbol{\theta}^{\text{best}}, \Sigma^{\text{best}})$
\For {$k=1$ to $M_w$}
    \State $d \gets 0$ (decrease counter).
    \State $s \gets 0$ (steps counter).
    \State $t_1 \gets s_1[k]$, $t_2 \gets s_2[k]$ (set to current time step).
    \State $m_{1,t_1} \gets b_{1}[k]$, $m_{2,t_2}\gets b_2[k]$ (set to best first moment vector/matrix).
    \State $v_{1,t_1} \gets c_1[k]$, $v_{2,t_2} \gets c_2[k]$ (set to best second moment vector/matrix).
    \While {ELBO for $\boldsymbol{\varphi}_k$ not converged}    
        \State $s \gets s+1$.
        \State $t_1 \gets t_1+1$.
        \State $h_{t_1} \gets -f_k(\boldsymbol{\theta}^{\text{curr}}, \Sigma^{\text{best}})$ (negate as maximisation).
        \State $m_{1,t_1} \gets \beta_1 m_{1,t_1 - 1} + (1- \beta_1) \cdot h_{t_1-1}$.
        \State $v_{1,t_1} \gets \beta_2 v_{1,t_1 - 1} + (1 - \beta_2)\cdot h_{t_1-1} \odot h_{t_1-1}$.
        \State $\hat{m}_{1,t_1} \gets m_{1,t_1} / (1 - \beta_1^{t_1})$.
        \State $\hat{v}_{1,t_1} \gets v_{1,t_1} / (1 - \beta_2^{t_1})$.
        \State $\boldsymbol{\theta}^{\text{curr}}[k] \gets \boldsymbol{\theta}^{\text{curr}}[k] - \eta_1 \cdot \hat{m}_{1,t_1} / (\sqrt{\hat{v}_{1,t_1}} + \epsilon)$.
        \State $\mathcal{L}^{\text{curr}} \gets \mathcal{L}(\boldsymbol{\theta}^{\text{curr}}, \Sigma^{\text{best}})$.
        \If {$\mathcal{L}^{\text{curr}} > \mathcal{L}^{\text{best}}$}
            \State $\mathcal{L}^{\text{best}} \gets \mathcal{L}^{\text{curr}}$.
            \State $\boldsymbol{\theta}^{\text{best}}[k] \gets \boldsymbol{\theta}^{\text{curr}}[k]$.
            \State $s_1[k] \gets t_1$.
            \State $b_1[k] \gets m_{1,t_1}$, $c_1[k] \gets v_{1,t_1}$.
        \Else 
            \State $d \gets d+1$.
            \If {$d > T$}
                \State Exit while loop.
            \EndIf
        \EndIf
        \If {$s > S$}
            \State Exit while loop.
        \EndIf
    \EndWhile
    \State $d \gets 0$.
    \State $s \gets 0$
    \While {ELBO for $\boldsymbol{\varphi}_k$ not converged}
        \State $s \gets s + 1$
        \State $t_2 \gets t_2+1$.
        \State $h_{t_2} \gets -g_k(\boldsymbol{\theta}^{\text{best}}, \Sigma^{\text{curr}})$ (negate as maximisation).
        \State $m_{2,t_2} \gets \beta_1 m_{2,t_2 - 1} + (1- \beta_1) \cdot h_{t_2-1}$.
        \State $v_{2,t_2} \gets \beta_2 v_{2,t_2 - 1} + (1 - \beta_2)\cdot h_{t_2-1} \odot h_{t_2-1}$.
        \State $\hat{m}_{2,t_2} \gets m_{2,t_2} / (1 - \beta_1^{t_2})$.
        \State $\hat{v}_{2,t_2} \gets v_{2,t_2} / (1 - \beta_2^{t_2})$.
        \State $\Sigma^{\text{curr}}[k] \gets \Sigma^{\text{curr}}[k] - \eta_2 \cdot \hat{m}_{1,t_1} / (\sqrt{\hat{v}_{1,t_1}} + \epsilon)$.
        \State $\mathcal{L}^{\text{curr}} \gets \mathcal{L}(\boldsymbol{\theta}^{\text{best}}, \Sigma^{\text{curr}})$.
        \If {$\mathcal{L}^{\text{curr}} > \mathcal{L}^{\text{best}}$}
            \State $\mathcal{L}^{\text{best}} \gets \mathcal{L}^{\text{curr}}$.
            \State $\Sigma^{\text{best}}[k] \gets \Sigma^{\text{curr}}[k]$.
            \State $s_2[k] \gets t_2$.
            \State $b_2[k] \gets m_{2,t_2}$, $c_2[k] \gets v_{2,t_2}$.
        \Else 
            \State $d \gets d+1$.
            \If {$d > T$}
                \State Exit while loop.
            \EndIf
        \EndIf
        \If {$s > S$}
            \State Exit while loop.
        \EndIf
    \EndWhile
    \State $\Tilde{\boldsymbol{\theta}}_{\varphi_k} \gets \boldsymbol{\theta}^{\text{best}}[k]$, $\widetilde{\Sigma}_{\varphi_k} \gets \Sigma^{\text{best}}[k]$.
\EndFor
\State Update $\hat{q}(\sigma_k^2)$ for $k \in M_w$ using \eqref{eqn:variational_nu}-\eqref{eqn:variational_omega}.
\State Update $\hat{q}(z_{\ell i})$ for $\ell \in [L]$ and $i \in \mathcal{V}$ using \eqref{eqn:variational_z}.
\State Updte $\hat{q}(w_i)$ for $i \in \mathcal{V}$ using \eqref{eqn:variational_w}.
\EndWhile
\end{algorithmic}
\label{alg:full_update_algorithm}
\end{breakablealgorithm}

\section{Additional figures}
\label{app:additional_figures}

\begin{figure}[!h]
\centering
\centering\includegraphics[width=\textwidth]{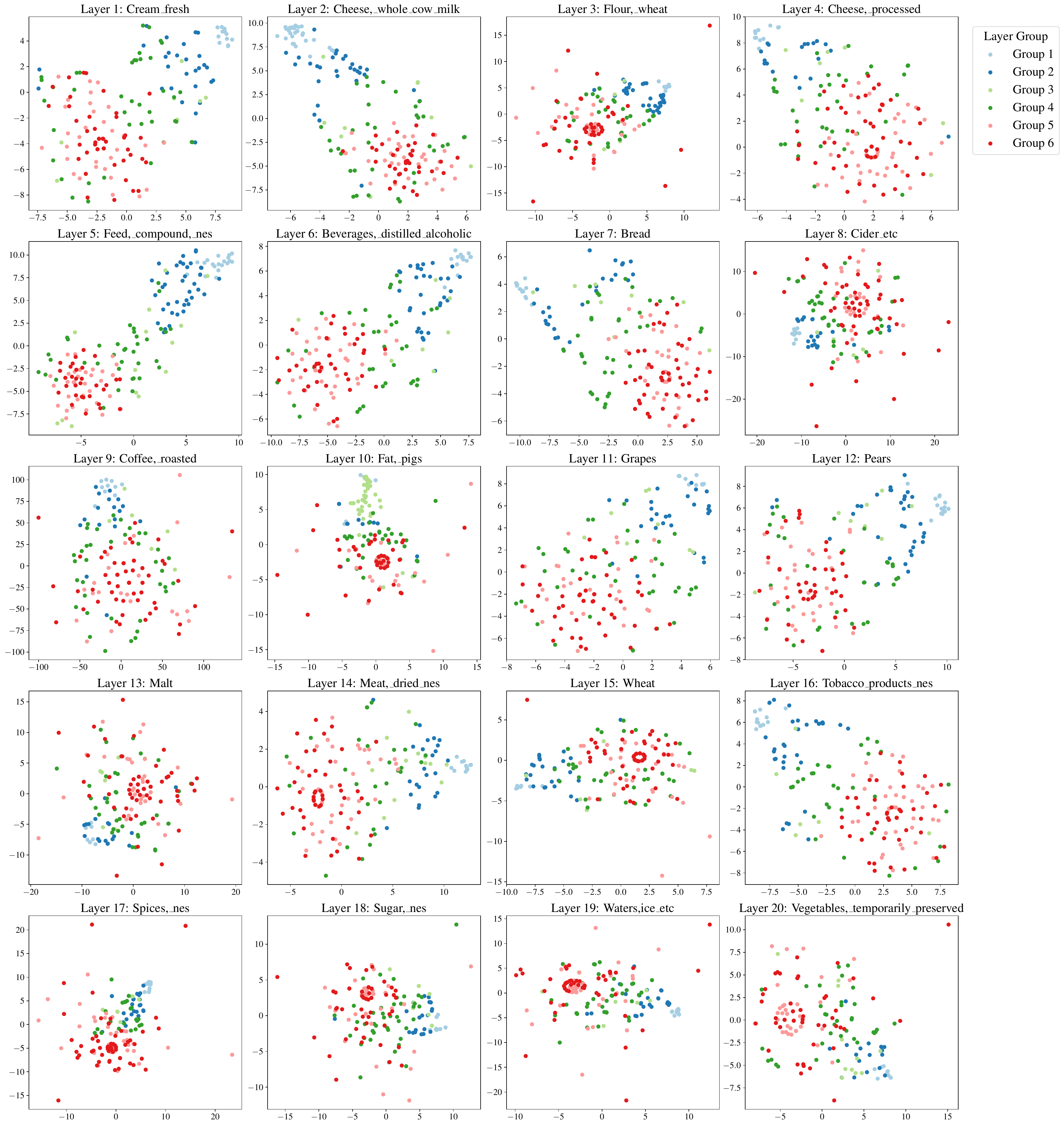}
    \caption{t-SNE for all layers in the network, with nodes coloured by their layer-level group.}
    \label{fig:tsne_all_layer}
\end{figure}
\end{appendices}

\end{document}